# Observation of metastable chiral domain walls in a topological magnet

**Authors:** Richen Xiong[1#], Chenxin Qin[1#], Zhaoyu Han[2], Nisarg Chadha[2], Qiang Gao[2], William Holtzmann[3], Weijie Li[3], Jiaqi Cai[3], Yi Guo[1], Weihanzhang Guo[1], Qi Chen[1], Samuel L. Brantly[1], Sam Bonkowsky[1], Chen Huang[1], Kenji Watanabe[4], Takashi Taniguchi[5], Andrea F. Young[1], Xiaodong Xu[3,6], Eslam Khalaf[2], Chenhao Jin[1]*

**Affiliations:**

[1]*Department of Physics, University of California at Santa Barbara, Santa Barbara, CA, USA.*

[2]*Department of Physics, Harvard University, Cambridge, MA, USA.*

[3]*Department of Physics, University of Washington, Seattle, WA, USA.*

[4]*Research Center for Electronic and Optical Materials, National Institute for Materials Science, Tsukuba, Japan.*

[5]*Research Center for Materials Nanoarchitectonics, National Institute for Materials Science, Tsukuba, Japan.*

[6]*Department of Materials Science, University of Washington, Seattle, WA, USA.*

# These authors contributed equally.

* Corresponding author. Email: jinchenhao@ucsb.edu

**Abstract:** The interplay between topology and correlation can give rise to exotic collective excitations[1–3]. The integer and fractional quantum anomalous Hall (QAH) magnets recently discovered in two-dimensional (2D) flatband systems[4–9] are predicted to host spin excitations distinct from those in conventional magnets[10–17]. Experimentally, nevertheless, these new excitations remain largely unexplored. Here we investigate spin-valley excitations in a twisted $MoTe_2$ moiré superlattice using resonant ultrafast pump-probe spectroscopy. We observe a metastable spin-valley excitation in the QAH magnet below $T \sim 3.7$ K that survives reverse magnetic field several times larger than the saturation field. The behavior of this excitation is sharply distinct from ordinary domain walls and magnons, indicating a new type of spin-valley textures unique to topological magnets. We propose that these textures are chiral domain walls with an in-plane winding of the pseudospin order parameter along the domain wall. Their metastability arises from the interplay between the topological winding in real space and the quantum geometry of the parent bands in momentum space through a universal mechanism. These chiral domain walls govern the nonequilibrium dynamics of QAH magnets and may play a central role in their stability. Our study highlights intrinsic quantum geometry effects on spin excitations in topological magnets; and provides key insights into the fundamental mechanism limiting stability of topological protection.

Excitations in a quantum system govern its response to external and internal perturbations. Topological magnets, such as integer and fractional quantum anomalous Hall (QAH) states, are expected to host exotic excitations such as fractionalized charge quasiparticles[9–13], magneto-rotons[14,15], topological magnons and domain walls[16,17]. These new excitations can be harnessed as information carriers for novel devices, including noise-resilient quantum computation[18]; and may give rise to dynamics and (in)stability mechanisms distinct from those in conventional magnets. 2D flatband systems recently emerged as an attractive platform for engineering topological magnetism[19,20]. While a plethora of intriguing ground states are reported[4–8,21–24], it has been challenging to directly probe the emergent excitations, particularly those decoupled from the charge sector[25]. Recently, optical switching of the spin-valley-polarized ground state has been demonstrated in twisted $MoTe_2$[26–28]. However, the stability of the switched states only reflects their ground state nature; while dynamics and excitations of the system remain unexplored. Therefore, central questions remain unanswered regarding how band topology and geometry influence the properties of the neutral excitations, and how these excitations affect the dynamics and stability of topological magnets[29,30].

Here we address these questions by directly probing spin-valley excitations in a twisted $MoTe_2$ moiré superlattice through resonant pump-probe spectroscopies. We observe multiple spin-valley excitations with distinct behaviors, which allow us to separate ordinary domains, magnons, and new excitations emerging in the QAH state. The emergent excitations are metastable and have a remarkably long lifetime of tens of microseconds. They govern the dynamics of the QAH state and appear in a temperature ($T$ < 3.7 K) and displacement field range substantially smaller than the spin-valley polarized (SVP) state. Interestingly, this region of phase diagram matches with the parameter space showing well quantized topological transport, suggesting a possible role of the new excitations in the stability of topological protection. We attribute these observations to chiral domain walls with a nontrivial winding texture, which gains stability against shrinking from the charge inhomogeneity induced by the mixed space-momentum Berry curvature.

**Emergent spin-valley excitations in the QAH state**

Fig. 1a and 1b show polarization resolved, doping dependent reflection contrast (RC) measurement of twisted $MoTe_2$ device D1 at out-of-plane magnetic field $B_z$ = 0, displacement field $D$ = 20 mV/nm, and nominal base temperature of 2.5 K (3.7° twist angle, the same device as device 1 in Ref.[4]). The right- and left-circularly polarized (RCP and LCP) light selectively probe optical responses from the K and K' valley, respectively[31]. Their prominent discrepancy in the moiré filling range of $-1.2 < \nu < -0.5$ indicates spontaneous time reversal symmetry breaking expected from a SVP ferromagnet, consistent with literature[4]. We first focus on the behavior at $\nu$ = -1, i.e., one hole per moiré period, which corresponds to a QAH state at low displacement field[4] (see Extended Data Fig. 1 for Streda formula measurement). Fig. 1c and 1d summarize the displacement field-dependent RC at $\nu$ = -1, which shows strong magneto-circular dichroism (MCD) over a large range. To pinpoint the region of spontaneous SVP, we measure magnetic hysteresis and summarize $\Delta MCD(B_z) = MCD_{up}(B_z) - MCD_{down}(B_z)$, where $MCD_{up}$ and $MCD_{down}$ correspond to MCD measured during forward and backward magnetic field scanning, respectively (see Extended Data Fig. 2). The hysteresis map (Fig. 1e) indicates that the system stays in an SVP state in the displacement field range -120 mV/nm < $D$ < 140 mV/nm (black dashed line), and the maximum saturation field is around 50 mT.

Next, we probe spin-valley excitations and dynamics using ultrafast resonant pump probe spectroscopies. We use a LCP femtosecond pump pulse at 1.123 eV (orange dashed line in Fig. 1a, FWHM is around 6 meV), in resonance with the attractive polaron peak. The resonant pump light ensures well-defined valley selection rules and selective injection of carriers into the K' valley[31] (see Extended Data Fig. 3 for comparison between resonant and non-resonant pump), which, after electron-hole recombination, relax into pure spin-valley polarization[32] (Fig. 1f). Meanwhile, we apply an external magnetic field $B_z$ = 100 mT, substantially larger than the saturation field, to maintain the equilibrium state polarized to the K valley. The optically injected spin-valley excitations will therefore eventually be flipped back, and the system returns to a fully SVP state (Fig. 1f). We capture such pure spin-valley relaxation process using a separate probe pulse, which detects pump induced spin-valley polarization change through the associated reflection change, $\Delta R/R$. To enhance the signal, we use spectrally narrow probe light (< 1 meV FWHM, see Methods) and set the center wavelength to be always at the repulsive polaron resonance (red dashed line in Fig. 1d).

Fig. 1g summarizes the displacement-field-dependent spin-valley dynamics at $\nu$ = -1. The probe light is configured to selectively detect spin-valley polarization (see Methods). We have also measured charge (population) dynamics, that is, electron-hole recombination, under identical pump using a different detection channel (Extended Data Fig. 4). The latter happens at a short time scale (< 100 ns) and is irrelevant to the spin-valley dynamics afterwards. Interestingly, we observe long-lived spin-valley excitations that barely decay within the full delay range of 10 μs, indicating a remarkable lifetime of at least tens of microseconds. In addition, it only appears at displacement field range of -60 mV/nm < $D$ < 40 mV/nm (white dashed line in Fig. 1g), which is substantially smaller than the range of SVP states (black dashed line in Fig.

1e).

**Separating different spin-valley excitations**

We perform systematic control experiments to elucidate the origin of the long-lived spin-valley excitations. The two well-documented spin excitations in conventional magnets are magnons and domains[33]. They are expected to show distinct excitation density dependence: magnons typically dominate at low excitation density, with a lifetime that decreases with density due to magnon-magnon interactions[34,35]. In contrast, magnons can merge into domains at high excitation density[36], whose lifetime increases with density since their energy cost follows a boundary law, and the energy landscape becomes flatter for larger domains. Fig. 2a and 2b compare the pump fluence dependent normalized spin-valley dynamics at $D$ = -6 mV/nm and 60 mV/nm, respectively (see Extended Data Fig. 5 for raw data). The results at $D$ = 60 mV/nm are well captured by domain dynamics, where the relaxation time increases with excitation density from a hundred nanosecond to one microsecond (Fig. 2g).

In contrast, two well separated components are observed at $D$ = -6 mV/nm (Fig. 2e, black squares). While the fast component is consistent with domain dynamics, the slow component shows exotic pump fluence dependence distinct from both magnons and domains. It is absent at low excitation density and appears abruptly above a threshold pump fluence F > 1.4 $\mu$J/cm$^2$ (Fig. 2f, black squares), opposite to magnons. In addition, it shows extremely slow relaxation regardless of the excitation density, at least two orders of magnitude slower than domain relaxation (Fig. 2g). Its unusual origin is further confirmed by measurement at higher temperature (Fig. 2c and 2d). While the domain dynamics remain largely unchanged between base temperature and 4.5 K (Fig. 2g), the slow component at $D$ = -6 mV/nm disappears completely (Fig. 2c). These observations are well-reproduced in two additional devices, D2 (3.6° twist angle) and D3 (3.7° twist angle) (see Methods).

We next explore evolution of this new excitation in the phase diagram. Fig. 3a and 3b summarize the temperature dependent spin-valley dynamics at $D$ = -6 mV/nm and 60 mV/nm, respectively. The corresponding equilibrium MCD hysteresis are shown in Fig. 3c and 3d. The slow component at $D$ = -6 mV/nm disappears sharply around 3.5 K (Fig. 3a), far below the $T_c$ of time-reversal symmetry breaking (> 12 K); while the dynamics at $D$ = 60 mV/nm only show slight changes in this temperature range. Above 3.7 K, the dynamics at the two displacement fields become indistinguishable, both insensitive to temperature until showing critical slowing down near $T_c$[37]. We further measure the effects of magnetic field $B_z$. Domain relaxation becomes faster under a larger positive $B_z$ (Fig. 3, g and h, red circles), which is expected from a more tilted free energy landscape that accelerates domain shrinking. In contrast, the emergent new excitations remain metastable even at a large $B_z$ = 200 mT (Fig. 3g, black squares). The dynamics at negative $B_z$ is always fast since the system is already fully polarized to the K' valley in equilibrium, and an LCP pump cannot generate more spin-valley polarization. We also measured carrier density dependent spin-valley dynamics at $D$ = 20 mV/nm (Fig. 3, i and j). The metastable excitations are only observed within a narrow filling range around one hole per moiré period at low temperature. To elucidate origin of spin-valley excitations away from $\nu$ = -1, we performed excitation density dependence at $\nu$ = -0.85 (Fig. 3k). The signal is well captured by ordinary domain dynamics, which does not show apparent changes at $\nu$ = -1 (Fig. 3j).

**Spin-valley texture stabilized by quantum geometry**

The emergent spin-valley excitation shows several remarkable properties distinct from conventional excitations in ferromagnets. It is metastable even under a reverse magnetic field several times larger than the saturation field and only appears in a parameter space substantially smaller than the SVP region. This again excludes magnons and ordinary domains, which are not metastable and should not show abrupt changes within the SVP state. Here we define metastability as a local minimum of the energy landscape separated from the global minimum by a finite activation barrier, which yields a slower relaxation than microscopic energy scales. In addition, the emergent excitation features a peculiar threshold behavior in density dependence and does not appear in the dilute limit (Fig. 2a), indicating its nature as a spin texture involving multiple spin flips. In conventional magnets, large spin textures are often long-lived because their relaxation involves mesoscopically many spins and a large barrier[38–40]. However, a large magnetic field would remove such a barrier and suppress the textures[40]. Our observation therefore indicates a new mechanism in topological magnets that can stabilize spin textures well beyond the saturation field.

In the following, we will propose such a mechanism. In an easy-axis ferromagnet, the spatial extent $d_0$ for any continuous spin texture, such as the width of a domain wall, is set by the competition between spin stiffness and easy-axis anisotropy. When $d_0$ is not microscopically small, the winding of the in-plane component along the wall is well-defined by an integer $N_w$ and endows topological protection. Our experiments show that ordinary domain walls have relatively short lifetime (Fig. 2). The metastable excitation therefore likely corresponds to a texture with non-zero winding number, which carries a chirality that defines the wall orientation and hence can be called a chiral domain wall (Fig. 4a). These chiral textures are generally stable against unwinding, but not against shrinking in conventional magnets.

In a topological magnet, in contrast, the band quantum geometry brings in new ingredients that further stabilize chiral domain walls. It is well known that skyrmions in quantum Hall ferromagnets are charged[41,42]. The chiral domain wall here, a ring-shaped skyrmion[40], leads to charge redistribution instead of charge accumulation due to the opposite Chern numbers inside and outside the domain wall. To the leading order allowed by the symmetries of the system, this effect can be described by a charge dipole density on the wall with a direction normal to it (Fig. 4b). The dipole density magnitude $p$ is proportional to the quantum geometric quantity $c_G$ of the electron bands, which is a weighted average of the second-Chern form over the four-dimensional manifold spanned by spin polarization and momentum[43–45] (see Supplementary Information). The dipole-dipole repulsion then yields an energy $\sim N_w^2 c_G^2/R$ for a chiral domain wall with radius $R$, which competes with the usual surface tension $\sim \sigma R$ and the Zeeman energy $\sim B_z R^2$ to determine an optimal domain size $R^*$ (Fig. 4c). The chiral domains are stabilized as long as $R^* > d_0$ so that no microscopic unwinding occurs, and can therefore survive magnetic field much greater than the saturation field. This picture also naturally explains the sensitive doping dependence of chiral domain walls, as the bound charge dipole is no longer stably definable for locally charge compressible systems. To quantitatively compare between experiment and theory, we map out the energy landscape of chiral domain walls by measuring their temperature dependence at different $B_z$. The extracted onset temperature $T^*$, at which chiral domain walls become unstable, reflects the $B_z$-dependent activation barrier height. $T^*$ decreases from 4.1 K at 80 mT to 3.5 K at 200 mT, corresponding to a slope $-dT^*/dB_z \approx 5$ K/T, in quantitative agreement with the theory estimated slope of 4 ~ 6 K/T (Fig. 4d, see Methods for details).

**Discussion and outlook**

Our results suggest a universal and previously unexplored effect of quantum geometry on spin textures in topological magnets with spin-contrasting Chern numbers, a limit distinct from the familiar quantum Hall ferromagnets with spin-degenerate Chern numbers. This gives rise to the metastability of chiral domain walls against microscopic relaxation. The intrinsic tie between quantum geometry and chiral domain walls should apply to a wide class of systems with spontaneously broken time-reversal symmetry and non-trivial quantum geometry, which provides an attractive tool for detecting quantum geometry of the electronic bands. For example, it has been theoretically proposed that topologically trivial and nontrivial SVP states coexist in $tMoTe_2$[46–49]. Indeed, the chiral domain walls are only stable within a subregion of SVP states (Fig. 1g), which can be consistent with the existence of multiple SVP states with distinct geometric properties.

The emergent metastable chiral domain walls govern the nonequilibrium dynamics of topological magnets. Recently, optical switching of QAH states is demonstrated in twisted $MoTe_2$ under resonant pump[26–28]. Our results provide a comprehensive view of switching dynamics: the excitons injected by circular pump relax into pure spin-valley polarization within 100 ns (Extended Data Fig. 4). At sufficiently large density, the flipped spin-valley becomes the majority, and the system switches to the opposite valley. At $D > 40$ mV/nm or $D < -60$ mV/nm, the flipped spin-valley stays as ordinary domains with a relatively short lifetime. It therefore takes a large pump intensity to reach the critical density. At -60 mV/nm $< D <$ 40 mV/nm, in contrast, chiral domain walls dominate, which have much longer lifetime and substantially reduce the required pump intensity. This explains distinct switching efficiency in the two cases reported in literature[26] (see Methods for more discussions). Meanwhile, the slow dynamics of chiral domain walls also limit the switching speed. Such fundamental tradeoff between efficiency and speed provides key guidelines for designing future devices based on topological magnets.

In addition, chiral domain walls are expected to play a central role in the stability of topological magnets, allowing us to potentially address several mysteries in previous studies. Magnetism in twisted $MoTe_2$ is observed to spontaneously fluctuate between 3 and 4 K around $\nu = -1$ and $D = 0$ (Extended Data Fig. 8 of Ref.[50]); and, counterintuitively, regain stability at higher temperature or large $D$-field where QAH state is expected to be less stable. Our results provide one plausible explanation: The chiral domain walls show incipient instability in the same parameter space, indicating their close energy competition with a different texture, such as ordinary domains. This can lead to strong fluctuations due to spontaneous transitions between the two configurations. At higher temperature or larger $D$-field, chiral domain walls are strongly unfavored in the competition, resulting in reduced fluctuations (see Methods for more discussions).

In summary, the stability of chiral domain walls could ultimately limit topological protection, similar to vortices limiting dissipationless transport in type II superconductors. This may explain why quantized transport is often observed at temperatures much lower than $T_c$ and single particle gap size[5,7], especially in devices with strong disorder (see Methods for more discussions on the role of disorder). Due to the intrinsic connection between quantum geometry and chiral domain wall stability, we expect similar mechanisms in general topological magnets, including fractional QAH states. Extended Data Fig. 7 investigates optical switching of the QAH ($\nu = -1$) and FQAH states ($\nu = -2/3$) in device D1 using a cryostat with lower base temperature of 1.6 K (see Methods). Indeed, similar peculiar temperature dependences are observed in both cases. At the QAH state, the switching is reliable below 3 K and above 4 K but becomes unstable in between due to critical instability of chiral domain walls. Such non-monotonic temperature dependence is even more prominent at $\nu$ = -2/3, where switching fails between 1.8 ~ 2.2 K but is reliable both below and above. This suggests a similar mechanism limiting the stability of FQAH states, i.e., chiral domain walls become unstable around 2 K. Due to the higher base temperature (2.5 K) of pump probe measurement, we cannot directly study chiral domain walls at $\nu$ = -2/3 (Extended Data Fig. 8). Future experiments at lower base temperature can further elucidate the dynamics and stability of FQAH states. Local probes such as Lorentz transmission electron microscopy will also provide valuable information by directly imaging the spatial textures of chiral domain walls.

**Reference:**

1. Stormer, H. L., Tsui, D. C. & Gossard, A. C. The fractional quantum Hall effect. *Rev. Mod. Phys.* **71**, S298–S305 (1999).

2. Savary, L. & Balents, L. Quantum spin liquids: a review. *Reports on Progress in Physics* **80**, 016502 (2017).

3. Keimer, B. & Moore, J. E. The physics of quantum materials. *Nat. Phys.* **13**, 1045–1055 (2017).

4. Cai, J. *et al.* Signatures of fractional quantum anomalous Hall states in twisted MoTe2. *Nature* **622**, 63–68 (2023).

5. Park, H. *et al.* Observation of fractionally quantized anomalous Hall effect. *Nature* **622**, 74–79 (2023).

6. Zeng, Y. *et al.* Thermodynamic evidence of fractional Chern insulator in moiré MoTe2. *Nature* **622**, 69–73 (2023).

7. Xu, F. *et al.* Observation of Integer and Fractional Quantum Anomalous Hall Effects in Twisted Bilayer MoTe2. *Phys. Rev. X* **13**, 031037 (2023).

8. Lu, Z. *et al.* Fractional quantum anomalous Hall effect in multilayer graphene. *Nature* **626**, 759–764 (2024).

9. Li, W. *et al.* Signatures of fractional charges via anyon–trions in twisted MoTe2. *Nature* **651**, 48–53 (2026).

10. Neupert, T., Santos, L., Chamon, C. & Mudry, C. Fractional quantum hall states at zero magnetic field. *Phys. Rev. Lett.* **106**, (2011).

11. Tang, E., Mei, J. W. & Wen, X. G. High-temperature fractional quantum hall states. *Phys. Rev. Lett.* **106**, (2011).

12. Sheng, D. N., Gu, Z. C., Sun, K. & Sheng, L. Fractional quantum Hall effect in the absence of Landau levels. *Nat. Commun.* **2**, (2011).

13. Reddy, A. P., Alsallom, F., Zhang, Y., Devakul, T. & Fu, L. Fractional quantum anomalous Hall states in twisted bilayer MoTe2 and WSe2. *Phys. Rev. B* **108**, (2023).

14. Kousa, B. M., Morales-Durán, N., Wolf, T. M. R., Khalaf, E. & MacDonald, A. H. Theory of Magnetoroton Bands in Moiré Materials. *Phys. Rev. Lett.* **135**, (2025).

15. Shen, X. *et al.* Magnetorotons in moiré fractional Chern insulators. *Phys. Rev. B* **113**, L081403 (2026).

16. Qiu, W. X. & Wu, F. Topological magnons and domain walls in twisted bilayer MoTe2. *Phys. Rev. B* **112**, (2025).

17. Xie, M. & Sarma, S. Das. Collective Spin Excitations in Correlated Moiré Chern Ferromagnets. http://arxiv.org/abs/2603.20370 (2026).

18. Nayak, C., Simon, S. H., Stern, A., Freedman, M. & Das Sarma, S. Non-Abelian anyons and topological quantum computation. *Rev. Mod. Phys.* **80**, 1083–1159 (2008).

19. Mak, K. F. & Shan, J. Semiconductor moiré materials. *Nat. Nanotechnol.* **17**, 686–695 (2022).

20. Cao, T., Fu, L., Ju, L., Xiao, D. & Xu, X. Fractional Quantum Anomalous Hall Effect. *The Annual Review of Condensed Matter Physics* **30**, 233–56 (2026).

21. Zang, J., Wang, J., Cano, J., Georges, A. & Millis, A. J. Dynamical Mean-Field Theory of Moiré Bilayer Transition Metal Dichalcogenides: Phase Diagram, Resistivity, and Quantum Criticality. *Phys. Rev. X* **12**, (2022).

22. Guo, Y. *et al.* Superconductivity in 5.0° twisted bilayer WSe2. *Nature* **637**, 839–845 (2025).

23. Xia, Y. *et al.* Superconductivity in twisted bilayer WSe2. *Nature* **637**, 833–838 (2025).

24. Xu, F. *et al.* Signatures of unconventional superconductivity near reentrant and fractional quantum anomalous Hall insulators. arXiv: 2504.06972 (2025).

25. Xiong, R. *et al.* Observation of propagating collective spin–valley modes in twisted WSe2. *Nat. Phys.* **22**, 877–883 (2026).

26. Holtzmann, W. *et al.* Optical control of integer and fractional Chern insulators. *Nature* **649**, 1147–1152 (2026).

27. Huber, O. *et al.* Optical control over topological Chern number in moiré materials. *Nature* **649**, 1153–1158 (2026).

28. Cai, X. *et al.* Optical switching of a moiré Chern ferromagnet. *Nature* **650**, 580–584 (2026).

29. Caio, M. D., Möller, G., Cooper, N. R. & Bhaseen, M. J. Topological marker currents in Chern insulators. *Nat. Phys.* **15**, 257–261 (2019).

30. Yu, J. *et al.* Quantum geometry in quantum materials. *NPJ Quantum Mater.* **10**, 101 (2025).

31. Xu, X., Yao, W., Xiao, D. & Heinz, T. F. Spin and pseudospins in layered transition metal dichalcogenides. *Nat. Phys.* **10**, 343–350 (2014).

32. Jin, C. *et al.* Ultrafast dynamics in van der Waals heterostructures. *Nat. Nanotechnol.* **13**, 994–1003 (2018).

33. Stöhr, J. & Siegmann, H. C. *Magnetism*. (Springer Berlin Heidelberg, Berlin, Heidelberg, 2006). doi:10.1007/978-3-540-30283-4.

34. Chumak, A. V., Serga, A. A. & Hillebrands, B. Magnon transistor for all-magnon data processing. *Nat. Commun.* **5**, 4700 (2014).

35. Chumak, A. V., Vasyuchka, V. I., Serga, A. A. & Hillebrands, B. Magnon spintronics. *Nat. Phys.* **11**, 453–461 (2015).

36. Mohseni, S. M. *et al.* Spin Torque–Generated Magnetic Droplet Solitons. *Science.* **339**, 1295–1298 (2013).

37. Jin, C. *et al.* Imaging and control of critical fluctuations in two-dimensional magnets. *Nat. Mater.* **19**, 1290–1294 (2020).

38. Nagaosa, N. & Tokura, Y. Topological properties and dynamics of magnetic skyrmions. *Nat. Nanotechnol.* **8**, 899–911 (2013).

39. Tserkovnyak, Y. & Xiao, J. Energy Storage via Topological Spin Textures. *Phys. Rev. Lett.* **121**, (2018).

40. Büttner, F., Lemesh, I. & Beach, G. S. D. Theory of isolated magnetic skyrmions: From fundamentals to room temperature applications. *Sci. Rep.* **8**,

4464 (2018).

41. Wu, F. & Das Sarma, S. Quantum geometry and stability of moiré flatband ferromagnetism. *Phys. Rev. B* **102**, (2020).

42. Khalaf, E. & Vishwanath, A. Baby skyrmions in Chern ferromagnets and topological mechanism for spin-polaron formation in twisted bilayer graphene. *Nat. Commun.* **13**, (2022).

43. Qi, X. L., Hughes, T. L. & Zhang, S. C. Topological field theory of time-reversal invariant insulators. *Phys. Rev. B Condens. Matter Mater. Phys.* **78**, (2008).

44. Xiao, D., Shi, J., Clougherty, D. P. & Niu, Q. Polarization and adiabatic pumping in inhomogeneous crystals. *Phys. Rev. Lett.* **102**, (2009).

45. Zhao, Y., Gao, Y. & Xiao, D. Electric polarization in inhomogeneous crystals. *Phys. Rev. B* **104**, (2021).

46. Zhang, X.-W. *et al.* Polarization-driven band topology evolution in twisted MoTe2 and WSe2. *Nat. Commun.* **15**, 4223 (2024).

47. Li, B., Qiu, W. X. & Wu, F. Electrically tuned topology and magnetism in twisted bilayer MoTe2 at νh = 1. *Phys. Rev. B* **109**, (2024).

48. Jia, Y. *et al.* Moiré fractional Chern insulators. I. First-principles calculations and continuum models of twisted bilayer MoTe2. *Phys. Rev. B* **109**, 205121 (2024).

49. Wang, T., Devakul, T., Zaletel, M. P. & Fu, L. Diverse magnetic orders and quantum anomalous Hall effect in twisted bilayer MoTe2 and WSe2. arXiv:2306.02501 (2023).

50. Park, H. *et al.* Observation of dissipationless fractional Chern insulator. *Nat. Phys.* **22**, 389–395 (2026).

51. Wang, L. *et al.* One-dimensional electrical contact to a two-dimensional material. *Science.* **342**, 614–617 (2013).

52. Jin, C. *et al.* Stripe phases in WSe2/WS2 moiré superlattices. *Nat. Mater.* **20**, 940–944 (2021).

53. Jin, C. *et al.* Imaging of pure spin-valley diffusion current in WS2 -WSe2 heterostructures. *Science.* **360**, 893–896 (2018).

54. Srivastava, A. *et al.* Optically active quantum dots in monolayer WSe 2. *Nat.*

*Nanotechnol.* **10**, 491–496 (2015).

55. Tongay, S. *et al.* Defects activated photoluminescence in two-dimensional semiconductors: Interplay between bound, charged, and free excitons. *Sci. Rep.* **3**, (2013).

56. Li, W. *et al.* Universal Magnetic Phases in Twisted Bilayer MoTe2. *Nano Lett.* **25**, 18044–18050 (2025).

57. Redekop, E. *et al.* Direct magnetic imaging of fractional Chern insulators in twisted MoTe2. *Nature* **635**, 584–589 (2024).

58. N. Chadha, Q. Gao, E. Khalaf, and Z. Han, To appear.

**Figure 1**

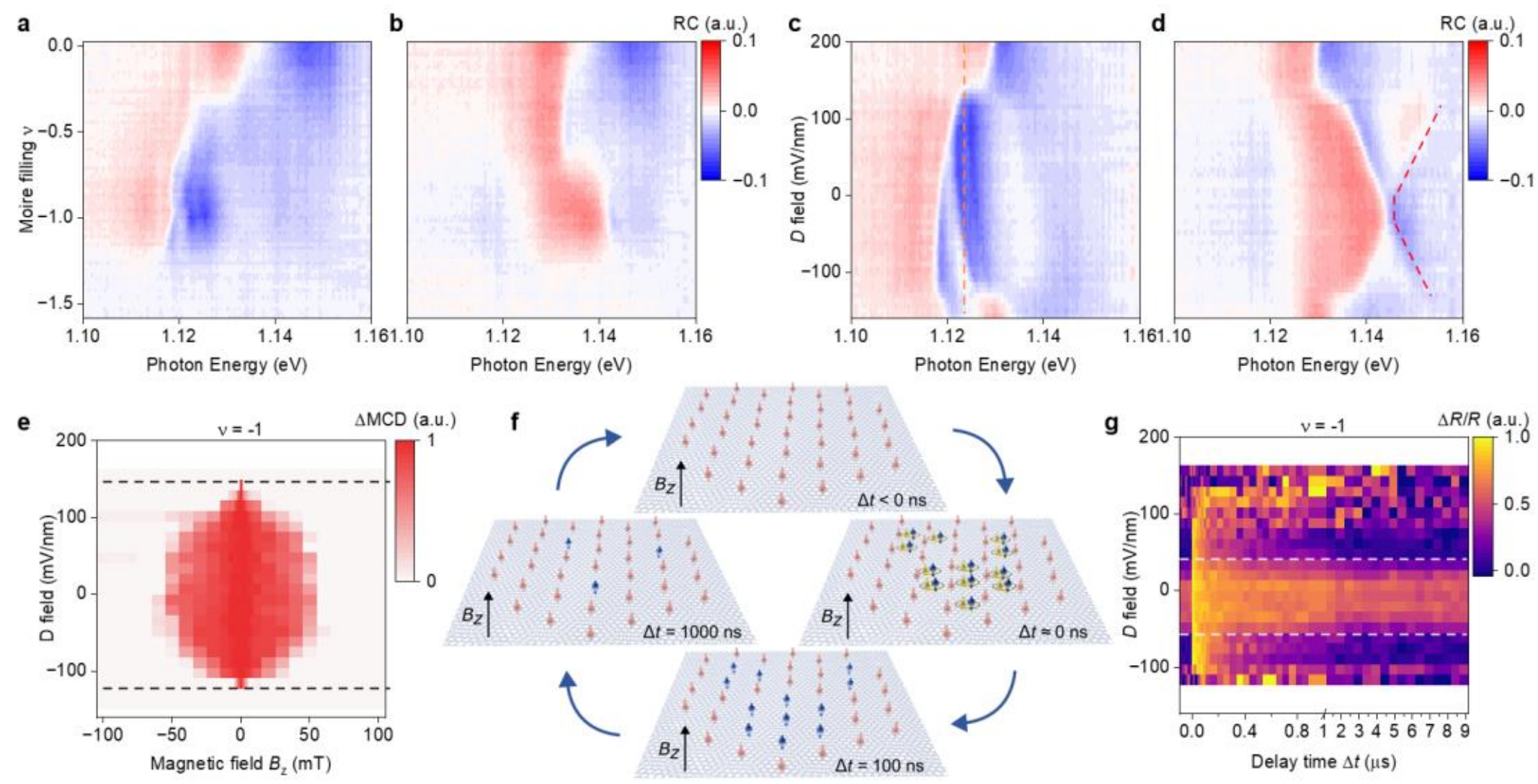

**Fig. 1: Emergent spin-valley excitations in the QAH state. a,b,** Polarization-resolved reflection contrast (RC) spectra of the twisted $MoTe_2$ device D1 as a function of moiré filling $v$ under LCP (**a**) and RCP (**b**) illumination at out-of-plane magnetic field $B_z = 0$ and displacement field $D$ = 20 mV/nm. **c,d,** $D$-field-dependent RC spectra at moiré filling $v$ = -1 under LCP (**c**) and RCP (**d**) illumination. Orange and red dashed lines label the photon energies of the pump and probe pulses used in resonant pump-probe measurements, respectively. **e,** Magnetic hysteresis map showing the difference in magneto-circular dichroism (ΔMCD) between forward and backward magnetic field sweeps at different $D$- fields. Black dashed lines outline the region of spontaneous spin-valley polarization (SVP), spanning $D$-fields from -120 mV/nm to 140 mV/nm. **f,** Schematic illustration of the resonant ultrafast pump-probe spectroscopy (see Methods). **g,** $D$-field-dependent normalized spin-valley dynamics at $v$ = -1 and $B_z$ = 100 mT. Long-lived metastable excitations emerge within a $D$-field range narrower than the SVP state (-60 mV/nm to 40 mV/nm, white dashed line). An axis break is applied at 1 μs to resolve both the fast and slow relaxation dynamics. The apparent fluctuations at high $D$-fields originate from reduced signal to noise ratio due to weak optical resonance. All measurements are taken at base temperature $T$ = 2.5 K.

## Figure 2

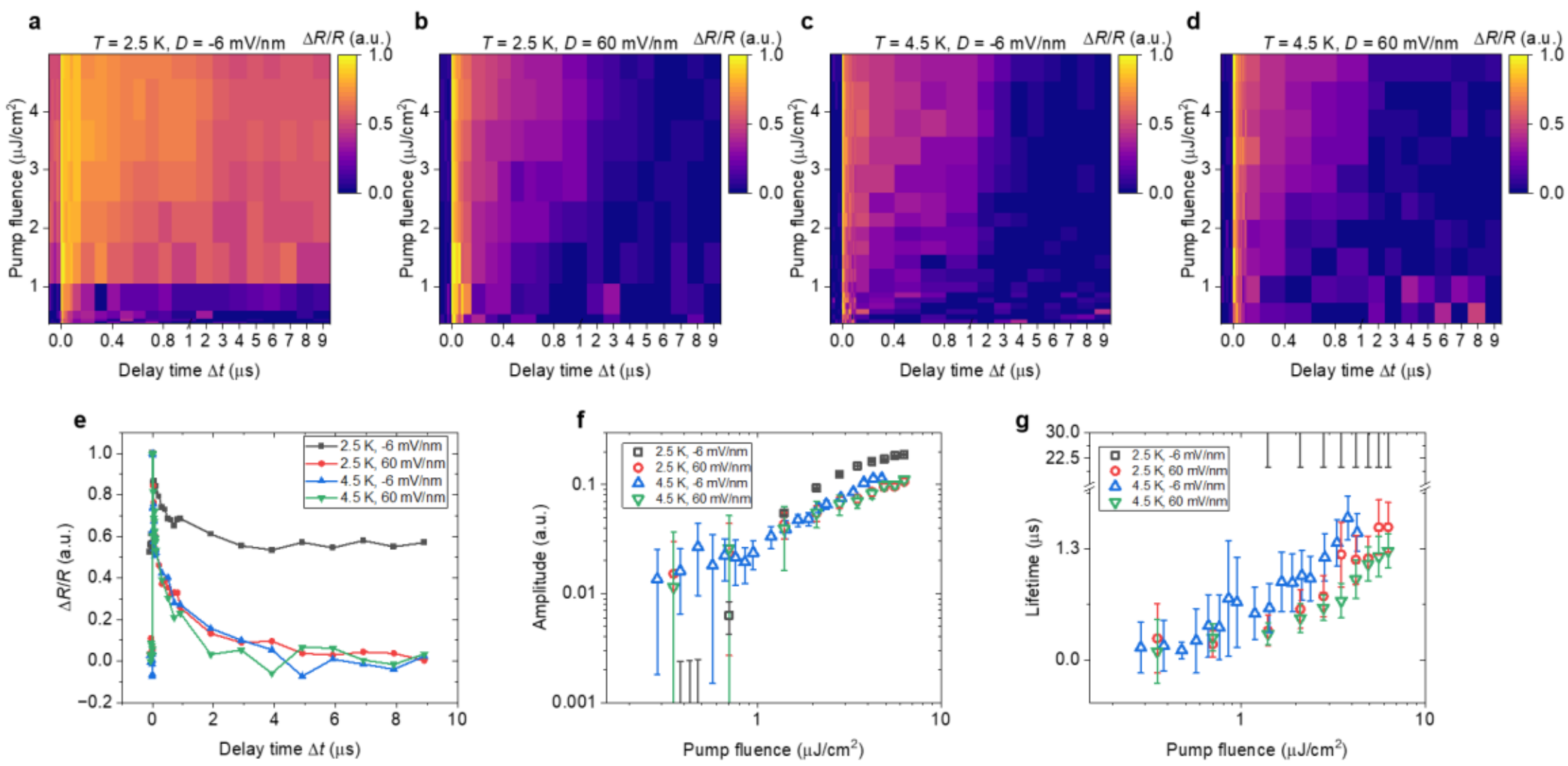

**Fig. 2: Separating different spin-valley excitations at $\nu$ = -1. a**,**b**, Pump fluence-dependent normalized spin-valley dynamics at $T$ = 2.5 K and $D$-field of -6 mV/nm (**a**) and 60 mV/nm (**b**), respectively. **c**,**d**, Same measurement as (**a, b**) at temperature of 4.5 K. The delay time axis breaks at 1 μs in (**a-d**). **e**, Representative normalized delay time traces at pump fluence F = 3.5 μJ/cm². Two well separated components are observed at $D$ = -6 mV/nm, $T$ = 2.5 K; and the slow component is fully suppressed at either larger $D$-field or higher temperature. The fast component is similar among all measurement configurations. **f**,**g,** Pump fluence-dependent amplitude (**f**) and lifetime (**g**) of spin-valley excitations under the four different configurations in (**a-d**) extracted from unnormalized data (see Extended Data Fig. 5). Only the slow component is shown for $D$ = -6 mV/nm, $T$ = 2.5 K as it dominates the signal. The fast component shows increasing lifetime at higher excitation density, indicating its origin from ordinary domain relaxation (see text). In contrast, the slow component shows unusual threshold behavior in amplitude (**f**) and remarkably long lifetime (**g**), distinct from both magnons and domains. All measurements are performed under $B_z$ = 100 mT. Error bars in (**f**) and (**g**) represent standard deviation of the fit. Only the lower bound of lifetime is estimated here for the slow component as it shows negligible decay in the entire delay range. An upper bound of hundreds of microseconds is obtained from a separate measurement (see Methods and Extended Data Fig. 9).

## Figure 3

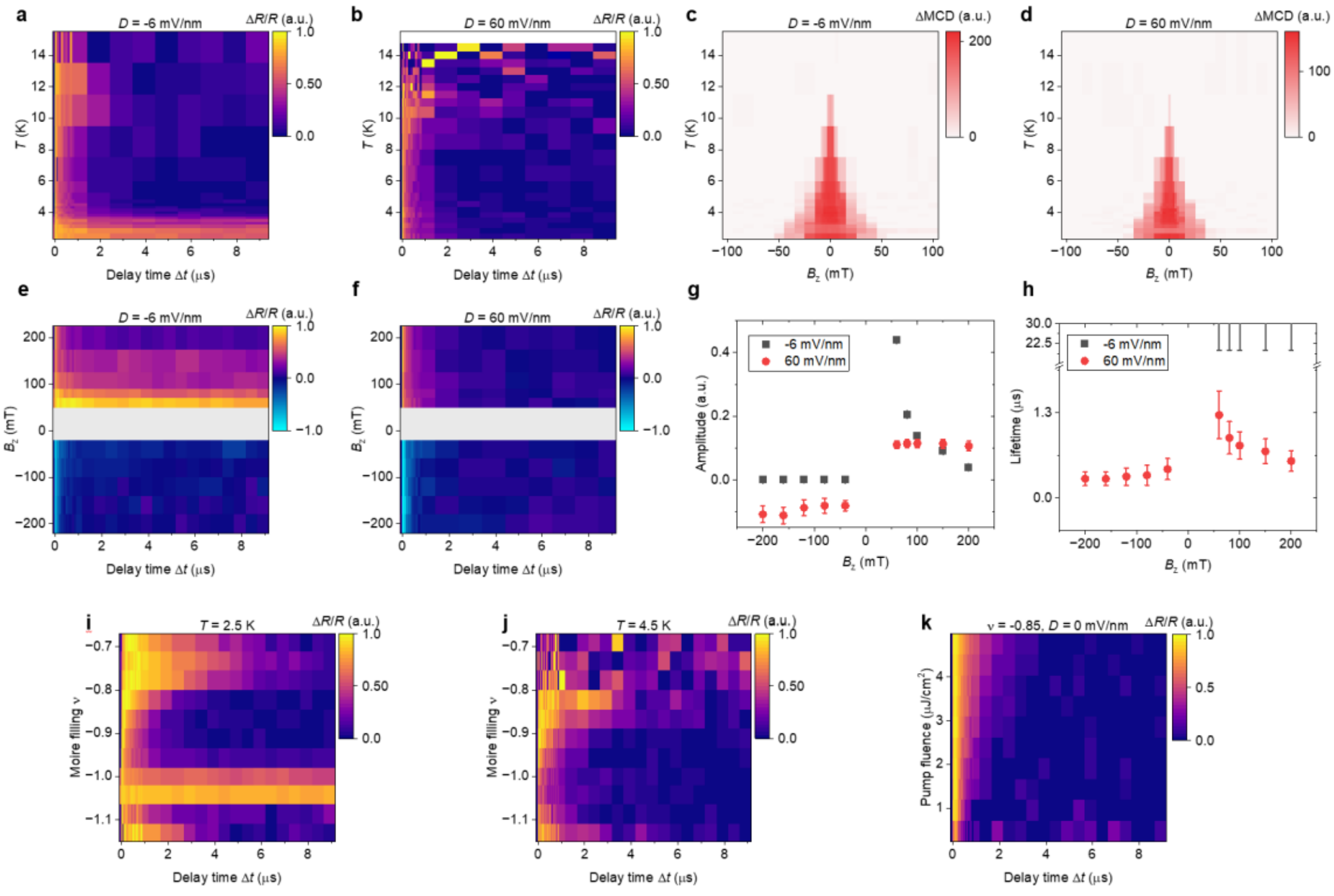

**Fig. 3: Evolution of the metastable spin-valley excitations. a**,**b**, Temperature-dependent normalized spin-valley dynamics measured at $D$ = -6 mV/nm (**a**) and 60 mV/nm (**b**), respectively. The slow component at $D$ = -6 mV/nm abruptly disappears around 3.5 K. **c**,**d**, Corresponding magnetic hysteresis maps as a function of temperature. **e-h**, Magnetic field-dependent normalized spin-valley dynamics measured at $D$ = -6 mV/nm (**e**) and 60 mV/nm (**f**), and the signal amplitude (**g**) and lifetime (**h**) extracted from unnormalized data. Domain relaxation becomes faster under larger reverse magnetic field. Remarkably, the emergent excitations remain metastable under reverse magnetic field of 200 mT, several times larger than the saturation field. Error bars in (**g**) and (**h**) represent standard deviation of the fit. **i**, Filling-dependent normalized spin-valley dynamics at $D$ = 20 mV/nm under $B_z$ twice the saturation field at each filling. The metastable excitations are observed in a narrow filling range around $\nu$ = -1. **j**, Same as (**i**) measured at 4.5 K. No apparent change in ordinary domain dynamics is observed across $\nu$ = -1. **k**, Pump-fluence-dependent normalized dynamics at $\nu$ = -0.85 under $B_z$ = 60 mT. The signal is fully accounted for by ordinary domain dynamics. Unless specified (as in (**k**)), all pump probe measurements here are performed with pump fluence of 3.5 μJ/cm².

**Figure 4**

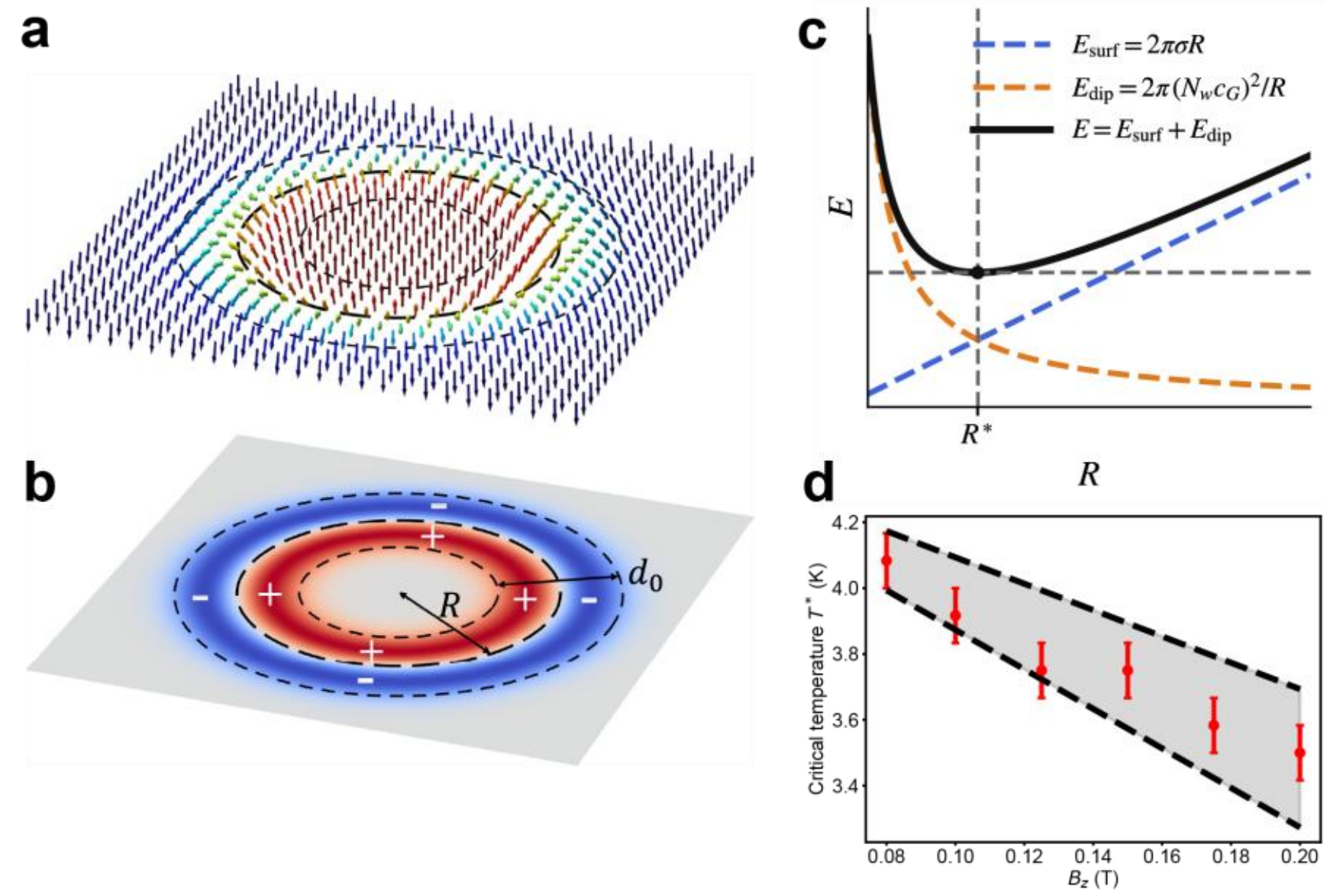

**Fig. 4: Chiral domain wall. a,** An illustration of the spin texture of a chiral domain wall. **b,** The charge density inhomogeneity induced by the spin texture, which is distributed around the wall with radius $R$ and width $d_0$. The dominant effect is characterized by a dipole density. **c,** Energy of a chiral domain wall as a function of $R$, including both surface-tension and dipolar-interaction contributions, showing a minimum at a finite radius $R^*$. **d,** Measured $B_z$-dependent onset temperature $T^*$ (red circles), at which the chiral domain walls become unstable. Error bars represent the uncertainty in determining $T^*$ from the temperature-dependent dynamics (Extended Data Fig. 10). Shaded region represents estimated slope of 4~6 K/T from microscopic calculation of quantum geometric responses (see Methods).

**Methods:**

**Sample preparation:** The tMoTe2 device 1 (3.7°) is the same device used in ref.[4]. The fabrication procedure was discussed in detail in ref.[4]. Device 2 (3.6°) and Device 3 (3.7°) were fabricated using a commercial bulk crystal (HQ graphene). The dual-gated twisted $MoTe_2$ devices were made by layer-by-layer dry transfer method[51] of van der Waals (vdW) materials as described in previous study[25]. Because $MoTe_2$ is air-sensitive, the exfoliation and assembly processes were performed within an argon-filled glovebox. The $MoTe_2$ flake was cut into two halves using an STM tip (P-50 PtIr) on the transfer station. The flakes were picked up by a polycarbonate (PC) thin film on a polydimethylsiloxane stamp with the following sequence: hBN, top-gate graphite, hBN, first half of the $MoTe_2$, second half of the $MoTe_2$ with a controlled twist, graphite contact, hBN and bottom-gate graphite. Finally, standard electron beam lithography and evaporation were used to pattern the contacts for wire bonding.

**Reflection contrast (RC) and photoluminescence (PL) measurements:** All measurements were performed in a closed-cycle cryostat (Quantum Design, Opticool) at a base temperature of 2.5 K unless specified. Keithley 2400 source meters were used to apply gate voltages. All light was collected by an InGaAs camera (C-RED 2, First Light Imaging) coupled with a spectrometer (SpectraPro HRS-300, Teledyne).
For RC measurements, probe light from a fiber-based supercontinuum laser (SC, YSL Photonics) was focused onto the sample through an objective (Olympus LCPLN20XIR, numerical aperture NA = 0.45), resulting in a beam diameter of approximately 1 μm on the sample with a power of approximately 2 nW. To probe the valley-resolved responses, a linear polarizer and a quarter-wave plate were used to generate RCP and LCP probe light, and the respective reflected intensities were collected. RC spectrum was computed as $RC = (R' - R)/R$, where R' is the intensity of the probe light reflected from the region of interest, and R is the reference spectrum obtained from heavily hole-doped states of the $tMoTe_2$. For zero-field measurements, the spin-valley polarization was first initialized by applying an out-of-plane magnetic field of 0.1 T, which was subsequently ramped back down to zero prior to data acquisition. For the PL measurements, the sample was excited using a 660 nm continuous-wave diode laser focused through an objective lens (BoliOptics LPlan 20x, NA = 0.4). The incident excitation power was maintained at 100 nW, and the PL spectra were integrated for 10 seconds per acquisition. A long-pass filter was positioned before the spectrometer to remove the excitation light.

**Static MCD and hysteresis measurements:** The MCD is measured through a polarization-resolved imaging technique as detailed in previous studies[37,52]. The probe light was produced from a SC and wavelength selected by home-built double monochromators to have a center photon energy around 1.15 eV with 0.5 meV full width at half max (FWHM). The energy of the probe light is fine-tuned to track the repulsive polaron peak of $tMoTe_2$ device. A pair of Glan–Thompson polarizers were used in combination to define and analyze the probe polarization, together with a broadband half-wave plate and quarter-wave plate positioned before the analyzer. Prior to interaction with the sample, the probe beam was linearly polarized and can be expressed as a coherent superposition of LCP and RCP components with equal amplitude and phase[37]:

$$E_{r0}\begin{pmatrix}1\\0\end{pmatrix} = \frac{E_{r0}}{2}\begin{pmatrix}1\\i\end{pmatrix} + \frac{E_{r0}}{2}\begin{pmatrix}1\\-i\end{pmatrix} \tag{1}$$

A finite spin-valley polarization $S_z$ in the system induces MCD, i.e., imbalance between LCP and RCP reflection. This leads to an electric field component perpendicular to the incident light, which can be sensitively picked up when the polarizers' transmission axes are close to orthogonal. Quantitatively, the MCD-induced relative reflection change is given by[37]:

$$\delta R \approx \frac{2}{\phi}\alpha \tag{2}$$

Here α denotes the MCD amplitude, and $\phi$ is the angle between the two polarizers ($\phi = 0$ corresponds to orthogonal). The MCD signal can be enhanced by orders of magnitude by choosing a small $\phi$. We take $\phi$ = 2° in our measurements.

**Ultrafast pump probe measurements:** Femtosecond pulses (1030 nm, 100 kHz, about 200 fs) were produced by seeding a regenerative amplifier with a mode-locked oscillator (Light Conversion, PHAROS). These pulses were subsequently guided into an optical parametric amplifier (Light Conversion, ORPHEUS) which produces pulse trains of 1.123 eV light as pump light, in resonance with the attractive polaron peak of tMoTe2 device. The probe light was produced from SC (100 kHz, 100 ps), triggered by an electronic pulse sourced from the regenerative amplifier. Relative delay between pump and probe pulses was achieved by feeding the electronic pulse through digital delay generator (Agilent 81110A, HP) prior to triggering the SC. The SC pulses were similarly wavelength selected as in static MCD measurements. Both pump and probe beams were independently expanded using imaging lenses before combining at a beamsplitter and impinging onto the sample. The reflected probe light was isolated from the pump light via a band pass filter and collected on the InGaAs camera. In order to obtain the pump-induced change in the probe reflection, the pump was modulated by an optical chopper at 7 Hz, which was synchronized with the detector. Comparison of images taken between pump "on" and "off" states allowed us to isolate relative reflection change as $\Delta R/R = (R_{pump\text{-}on} - R_{pump\text{-}off})/R_{pump\text{-}off}$. The typical light intensity we use is 0.1 nW/$\mu$m$^2$ for probe light and 4 nW/$\mu$m$^2$ for pump light. Because the magnetic state of the system is highly susceptible to light, we carefully minimize the probe intensity to ensure that probe light does not alter sample magnetization.

**Separating charge and spin-valley excitations**: We directly separate charge and spin-valley excitations through different configurations of the probe beam[37,52]. As shown in Eq. 2, under near-cross configuration, the pump-induced reflectivity change, ΔR/R, is proportional to the pump-induced MCD change of the sample, which selectively probes spin-valley excitations. Conversely, in parallel configuration ($\phi = 90°$), the probe light measures the polarization-unresolved reflectivity of the sample, which captures the total responses of the two spin-valleys and therefore the charge (population) excitations. This allows us to obtain a complete picture of nonequilibrium dynamics. As illustrated in Fig. 1f, the K' excitons initially created by the resonant pump carry both charge and spin-valley excitations. Because the system has much larger hole density in K valley (spin up), pump-injected electrons have a high probability to recombine with K (spin up) holes, resulting in pure spin-valley flips after electron-hole recombination. We directly capture the electron-hole recombination process using the parallel detection channel (Extended Data Fig. 4a and 4b), which shows a relatively short, pump fluence-independent lifetime of < 100 ns. Therefore, all signals afterwards correspond to pure spin-valley excitations decoupled from charges. This is further confirmed by $\phi$-dependence. According to Eq. 2, signals from spin-valley excitations would change signs when reversing the sign of $\phi$, while signals from charge (population) excitations are independent of $\phi$. Indeed, all signals after 100 ns change sign with $\phi$ (Extended Data Fig. 4c), confirming their origin from pure spin-valley excitations.

**Determination of carrier density and displacement field:** The charge carrier density $n$ and the out-of-plane displacement field $D$ are independently controlled by the top- and bottom-gate voltages ($V_{tg}$ and $V_{bg}$). We calculate these parameters from the applied gate voltages using a standard parallel-plate capacitor model: $n = (V_{tg}C_{tg} + V_{bg}C_{bg})/e - n_0$ and $D/\varepsilon_0 = (V_{tg}C_{tg} - V_{bg}C_{bg})/2\varepsilon_0 - D_0$, where $C_{tg}$ and $C_{bg}$ are the geometric capacitance of top and bottom gate, $e$ is the electron charge and $\varepsilon_0$ is the vacuum permittivity. The offset carrier density $n_0$ is derived by aligning the electrostatic model with the well-defined $\nu$ = -1 and $\nu$ = -2/3 states observed in the dual-gate MCD (Extended Data Fig. 2a). The offset displacement field $D_0$ is determined from the symmetric axis of the dual-gate MCD map (Extended Data Fig. 2a). Finally, we calculated the twist angle based on the carrier density of the moiré one filling.

**Optical control of magnetization:** Optical control measurements in Ext. Data Fig. 7

a-f were performed in a closed-loop magneto-optical cryostat (attoDRY2100) at a base temperature of 1.6 K. An external magnetic field (-0.2T) is used to initialize the system into a full SVP state with positive degree of circular polarization (DOCP) in PL. Afterwards, the magnetic field is removed, and a pump beam switches the system to the opposite magnetization. The pump beam was generated by a supercontinuum source (NKT SuperK Fianium FIU-15; 78 MHz repetition rate, 35 ps pulse width) and spectrally filtered to the trion resonance using a home-built single-grating double-subtractive monochromator, yielding a linewidth of ~1 meV. The pump polarization was controlled by a linear polarizer followed by a quarter-wave plate to produce either $\sigma^+$ or $\sigma^-$ excitation, and the beam was focused onto the sample to a spot size of approximately 2 μm. After optical switching, the pump beam was turned off, and the SVP state was read out by measuring helicity-resolved PL under excitation with a linearly polarized 632.8 nm HeNe probe laser. The emitted PL passed through a quarter-wave plate and a linear polarizer to select either $\sigma^+$ or $\sigma^-$ polarization and was then directed into a spectrometer (Teledyne Princeton Instruments SpectraPro HRS-500). The PL was dispersed by a 600 grooves/mm diffraction grating blazed at 1 μm and detected using a liquid-nitrogen-cooled InGaAs photodiode array (PyLoN-IR 1.7). The helicity resolved PL spectra was acquired using 5 nW HeNe laser excitation and 120 s integration time.

Optical control measurements in Ext. Data Fig. 7 g-h were performed under the same configurations as the pump probe measurements in the main text, except without external magnetic field or pump-probe synchronization. The device is exposed to the pump light for 1 second. Afterwards, the probe light measures the spin-valley polarization of the device 6 times, each integrating for 3 seconds, to probe stability of the switched state. The pump light polarization alternates between LCP and RCP to repeatedly switch the device between the two degenerate spin-valley polarized ground states.

**Connection between dynamics and optical switching:** In the optical switching experiments (Ref.[26–28]), the states after switching are the ground states of the system, which, by definition, have an indefinitely long lifetime and do not provide information about excitations or dynamics. Our measurement uses a fundamentally different configuration. First, an external magnetic field resets the system to its initial state after each pump pulse. As a result, the contribution of the ground state cancels in the pump probe signal, isolating the dynamics of the spin-valley excitations. Second, we utilize ultrafast synchronized pump and probe lights, which enable direct investigation of these dynamics. We demonstrate the distinct information from the dynamics and switching measurements by directly comparing them under the same pump conditions. As shown in Extended Data Fig. 7g, h, magnetism in device D1 can be reliably switched at all *D*-fields and remains stable for at least 20 seconds after switching, as expected from the indefinite lifetime of a ground state. In contrast, the spin-valley relaxation time changes dramatically with *D*-field and can be as short as hundreds of nanoseconds (Fig. 2g). This comparison confirms that our measurement isolates the dynamics of spin-valley excitations without a ground state contribution, whereas the switching measurements probe only the switched ground state. The long-lived signal we observe is therefore not complicated by ground state switching and instead reflects emergent spin-valley excitations in the QAH state.

Meanwhile, the spin-valley dynamics measured here provide key insight into the switching process: while they are irrelevant to the stability of the switched ground state, they govern the switching efficiency. Intuitively, successful switching requires the pump light flip majority of holes in the system to the opposite spin-valley; and the maximum spin-valley flips depend on both the pump strength and the spin-valley relaxation time. In the continuous wave (CW) limit, the density of steady state spin-valley flips induced by the pump light is proportional to the spin-valley lifetime. Therefore, a longer spin-valley lifetime will lead to lower pump light intensity needed to flip majority of spin-valleys and increase the switching efficiency. On the other hand, in the limit of single pump pulse, the maximum density of spin-valley flips only depends on the pump pulse fluence but not the spin-valley lifetime. Therefore, the spin-valley lifetime does not affect the switching efficiency in this limit. In practice, the pump light has a finite repetition rate and falls within the two limits depending on the comparison between the repetition rate and the spin-valley lifetime. In Ref.[26], pump light has a relatively high repetition rate of 78 MHz, which is closer to the CW regime. Therefore, the switching efficiency is found to be considerably higher at lower *D*-field due to the much longer spin-valley lifetime revealed by our work. Our measurements use a much lower pump repetition rate of 100 kHz, which is closer to the single-pulse regime. Therefore, the switching efficiency should be much less sensitive to *D*-field, consistent with our observations.

**Estimation of the spin flip density:** The relaxation between valley-polarized excitons into pure spin-valley polarization is a universal process in two-dimensional transition metal dichalcogenides and heterostructures[25,32,53]. The key mechanism is that electrons within optically excited excitons may recombine with any hole in the system. If the system is initially hole-doped, electrostatically doped holes exist in both valleys. When the density of these holes is much larger than optically injected holes, the total hole density is approximately the same between the two valleys. Therefore, optically injected electrons have similar probability to recombine with holes in both valleys, which do not reduce net valley polarization. After such recombination, the system will have more holes in one valley but less holes in the other valley with zero net charge, i.e., pure spin-valley polarization. In the present system with spin-valley polarized ground state, all electrostatically doped holes stay in one valley (e.g. K valley). When the density of these holes is much larger than optically injected holes, optically injected electrons will mostly recombine with holes in the K valley due to the much higher hole density. Therefore, each K' exciton will inject net spin-valley polarization of two – by flipping one hole from K to K' – instead of one as in the previous case[32]. A more qualitative difference is that, in the present case, even an unpolarized pump light can inject net spin-valley polarization, because the system itself already breaks time-reversal symmetry. In a time-reversal-symmetric system, by contrast, an unpolarized pump light cannot inject net spin-valley polarization.

We calibrate the density of pump-induced spin-valley flips through two independent ways. First, each valley polarized exciton results in roughly one spin-valley flip[25,53]. Assuming a pump fluence of 5.4 μJ/cm² and absorption of 5%, each pulse creates an exciton density of $1.5\times10^{12}$ cm$^{-2}$ and therefore similar density in spin-valley flips. Alternatively, we find that the maximum spin-valley polarization signal starts to saturate above pump fluence of 5.4 μJ/cm², which should correspond to a substantial fraction of the holes in the system are spin-valley flipped. Using the total hole density of $3.8\times10^{12}$ cm$^{-2}$ at $\nu = -1$ in 3.7-degree twisted $MoTe_2$, we estimate the density of spin-valley flips to be $1{\sim}2\times10^{12}$ cm$^{-2}$ at pump fluence of 5.4 μJ/cm², consistent with the first calibration method. Based on these estimations, the long-lived spin-valley excitations are observed both in the perturbation regime (~10% flips) and the switching regime (~50% flips), see Fig. 2a in the main text. This confirms that our measurements reflect intrinsic dynamics of spin-valley excitations and are not affected by ground state switching.

**Estimation of the chiral domain wall lifetime:** The dynamics with (without) the emergent long-lived spin-valley excitations are fitted using standard bi-exponential (single exponential) model

$$A(t) = A_1 e^{-\frac{t}{\tau_1}} + A_2 e^{-\frac{t}{\tau_2}} \quad (3)$$

where $A$ is total signal, $t$ is the pump probe delay, $A_1, A_2$ and $\tau_1, \tau_2$ are the amplitude and lifetime of the two components, respectively. $A_2 = 0$ in the case of single exponential fitting. Each experimental data point was given the same weight in the fitting, and no offsets or background subtractions were introduced. One example fitting is shown in Extended Data Fig. 9a. Error bars in the main figures represent the standard deviation of the fit. In principle, Eq. 3 still works even if the lifetime is longer than the period of pump pulses. Assuming each pump pulse generates signal independently, we have:

$$A(t) = A_1 \sum_{m=0}^{N-1} exp\left[-\frac{t+mT}{\tau_1}\right] + A_2 \sum_{m=0}^{N-1} exp\left[-\frac{t+mT}{\tau_2}\right]$$
$$= A_1 e^{-\frac{t}{\tau_1}}\left(\sum_{m=0}^{N-1} exp\left[-\frac{mT}{\tau_1}\right]\right) + A_2 e^{-\frac{t}{\tau_2}}\left(\sum_{m=0}^{N-1} exp\left[-\frac{mT}{\tau_2}\right]\right) \quad (4)$$

where $T$ is the period of pump pulses, and $N$ is the total number of pump pulses before time $t$. Therefore, while the amplitude of the components can increase due to accumulation over multiple laser cycles, their dynamics are not affected by such accumulation. Extended Data Fig. 9b simulates the measured dynamics of a single component for different $\tau$ with $N = 50$. In all cases, the dynamics follow simple exponential decay. For comparison, we also plot the experimental dynamics at $t > 2$ μs (black symbols in Extended Data Fig. 9b), which indicate a lifetime of at least 50 μs. An upper bound cannot be extracted accurately by this method, because the difference between the simulated curves becomes very small when $\tau \gg T$.

We estimate the upper bound of $\tau$ through a separate measurement. Essentially, the pump probe signal is the AC response of the system to the modulated pump light, obtained through lock-in detection. If the pump modulation (through an optical chopper in our measurement) is much faster than the intrinsic dynamics, the system will not be able to respond. Following this idea, we measured the pump-induced spin-valley polarization at a fixed delay $\Delta t = 9.99$ μs while sweeping the pump modulation (chopper) frequency and keeping all other conditions identical (Extended Data Fig. 9c). The signal shows a negligible decrease up to a modulation frequency of 300 Hz, indicating dynamics faster than 1 kHz (the red curve is the standard response of a 1 kHz bandwidth). This sets an upper limit of a few hundred μs on the lifetime.

**Role of disorder:** Native, quenched disorder should not have a qualitative effect on the observed pump probe dynamics. The injection of spin-valley flips is largely unaffected by disorder, as the creation and recombination of electron-hole pairs happens at a high energy scale of interband transition. Once the spin-valley flips are excited by the pump light, these "optically created disorders" have much higher density than quenched disorders and should dominate the signal. Responses from quenched disorders can also be excluded by pump fluence dependence (Fig. 2a in the main text). Quenched disorders typically play a more dominant role at low excitation density but will be saturated at higher density[54,55]. Experimentally, however, the long-lived spin-valley excitations do not appear at low excitation density and only emerge above a threshold density, which is opposite to the behavior of quenched disorders. To further exclude complications from disorder, we performed additional measurements on a third device D3 (3.7° twist angle), summarized in Extended Data Fig. 6i-o. This device shows broader linewidths in reflection contrast, indicating a higher level of quenched disorder, yet all major observations are well reproduced. This confirms that our results capture the intrinsic response of a topological magnet and are not qualitatively affected by disorder.

Meanwhile, disorder and chiral domain walls together provide one plausible mechanism behind the spontaneous magnetization fluctuations between 3 to 4 K[4,50,56]. Due to spin-valley locking in the current system, magnetization fluctuations necessarily originate from spin-valley excitations. Their unusual temperature dependence requires the dynamics of such spin-valley excitations to change substantially between 3 and 4 K, which is well below $T_c$. Indeed, we observe that chiral domain walls become unstable in this temperature range. Microscopically, when the size of a chiral domain is smaller than the domain wall width, it becomes indistinguishable from an ordinary domain and can transition into the latter through microscopic unwinding. Therefore, in the clean limit, once thermal energy is sufficiently high to overcome the activation barrier, chiral domain walls have finite probability to evolve into an ordinary domains and decay rapidly afterwards. This results in one-way decay of chiral domain walls, as observed in our pump probe measurements. On the other hand, disorders in the system can pin magnetic domains and prevent them from complete annihilation. Phenomenologically, this adds a further term to the free energy that keeps the domain size finite, thereby creating a double well in the energy landscape. With sufficient thermal activation energy, the system can spontaneously transition between the two wells, resulting in magnetic fluctuations. Experimentally, macroscopic-scale magnetic fluctuations in the QAH state only appear in $MoTe_2$ devices with a relatively high disorder level[4,50,56], consistent with our picture.

**Quantitative comparison between experiment and theory:** The exclusive emergence of the metastable excitations at low $D$-field indicates qualitative differences between the states at low and high $D$-fields, which, however, are both insulators with a large charge gap[6,57] and have similar spin parameters (Fig. 3). Meanwhile, it is widely proposed that $D$-field should lead to a topological transition at $\nu$ = -1 in the present system[46,47,49]. Therefore, the most probable qualitative difference between the two states is their different topology. Indeed, recent microscopic Hartree-Fock calculations show that 3.7-degree-twisted MoTe2 transitions from a QAH state to a topologically trivial spin-valley polarized state upon increasing $D$-field. The quantum geometry quantity $c_G$, which determines the stability of chiral domain walls, decreases abruptly at the first-order topological transition[58]. Experimentally, the metastable spin-valley excitations only emerge within a subregion – about half in the $D$-field range – of the spin-valley polarized ferromagnet and disappear abruptly upon increasing $D$-field, which matches quantitatively with the first order topological transition from theory.

Our work highlights the central role of a high-order quantum geometry quantity, $c_G$, in topological magnets. Intuitively, it probes how the entire occupied Bloch subspace evolves as the spin-valley order parameter rotates, rather than only the Berry curvature or quantum metric evaluated at a fixed magnetic configuration; and therefore corresponds to a higher order quantum geometry quantity. $c_G$ controls the interplay between chiral spin texture and band quantum geometry, giving rise to dipole accumulation on chiral domain walls and determining their metastability. Using spin parameters from literature[16] and calculated $c_G$, theory predicts a slope of $-\mathrm{d}T^*/\mathrm{d}B_z$ around 4~6 K/T. The quantitative consistency between the experiment and theory (Fig. 4d) further confirms the observed metastable spin-valley excitations from chiral domain walls. A complete list of parameters used in theoretical calculations and the full derivation are provided in the supplementary information.

**Role of chiral edge mode:** The chiral edge modes and the quantum-geometric dipole are, respectively, longitudinal and transverse charge responses to the same topological domain wall. A domain wall separating regions whose Chern number differs by two should host two chiral modes traversing the bulk gap. In the smooth-wall limit, these modes can propagate along the wall but are localized along the transverse direction. Meanwhile, the quantum-geometric response describes the transverse redistribution of charge, which generates a radial dipole moment without net charge. Because the one-dimensional chiral edge modes are only mobile along the wall, they will not affect the transverse charge distribution in the leading order. In contrast, mobile carriers in a compressible bulk are two-dimensional and can move both along and perpendicular to the domain wall, which can efficiently neutralize the texture-induced transverse charge redistribution. Therefore, the stability of chiral domain walls is sensitive to the two-dimensional bulk incompressibility, but not to the one-dimensional chiral edge modes.

**Supplementary Information:** Phenomenological theory of the metastable spin excitations.

**Data availability:** All data and code are available from the corresponding authors upon reasonable request.

**Competing interests:** The authors declare no competing interests.

**Extended Data Figures:**

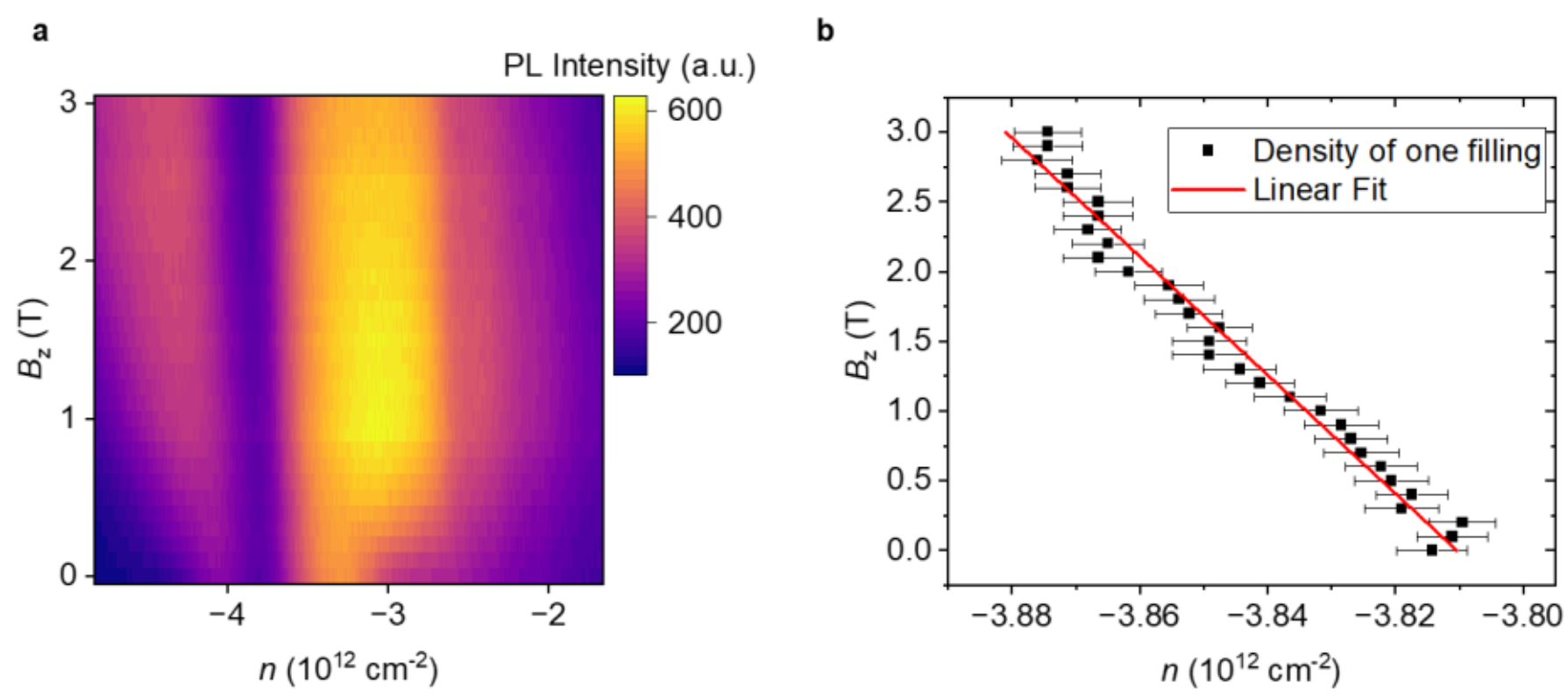

**Extended Data Fig. 1: Streda formula measurements for device D1. a,** Integrated photoluminescence (PL) intensity map as a function of out-of-plane magnetic field $B_z$ and carrier density $n$ at $D = 0$ mV/nm. **b,** Extracted density versus magnetic field $B_z$ for the incompressible state around one moiré filling. The linear fit yields a slope corresponding to Chern number C = 0.973 $\pm$ 0.028, consistent with a QAH state.

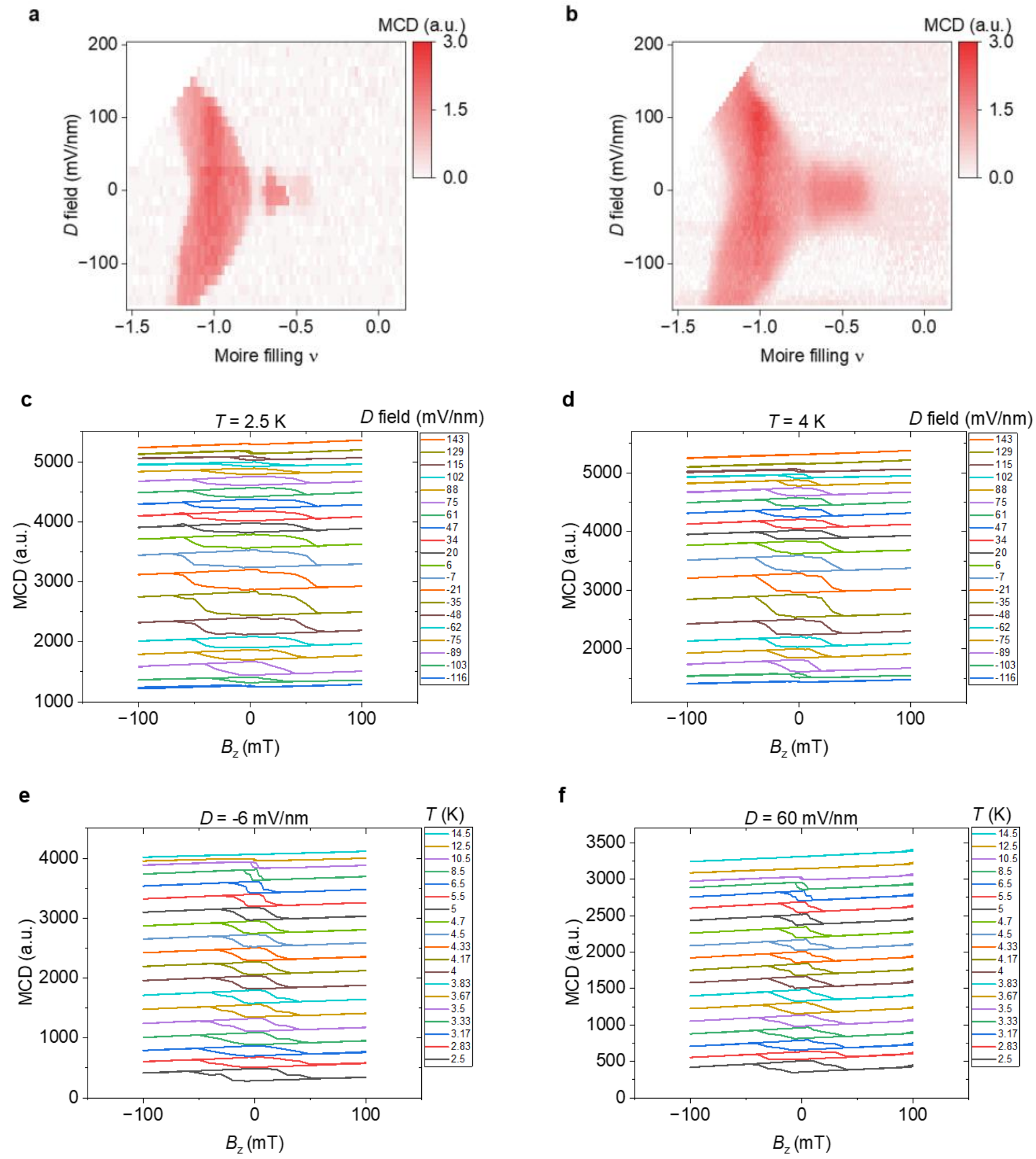

**Extended Data Fig. 2: Dual-gate MCD maps and magnetic hysteresis of device D1. a,b,** MCD as a function of out-of-plane displacement field $D$ and moiré filling $\nu$ under external magnetic field $B_z = 0$ (**a**) and $B_z = 1$ T (**b**) at $T = 2.5$ K. **c,d,** MCD hysteresis loops at $\nu = -1$ and different $D$-fields for temperatures of $T = 2.5$ K (**c**) and $T = 4$ K (**d**). **e,f,** MCD hysteresis loops at $\nu = -1$ and different temperatures for $D = -6$ mV/nm (**e**) and $D = 60$ mV/nm (**f**). Each curve in **c-f** is shifted up successively for visual clarity.

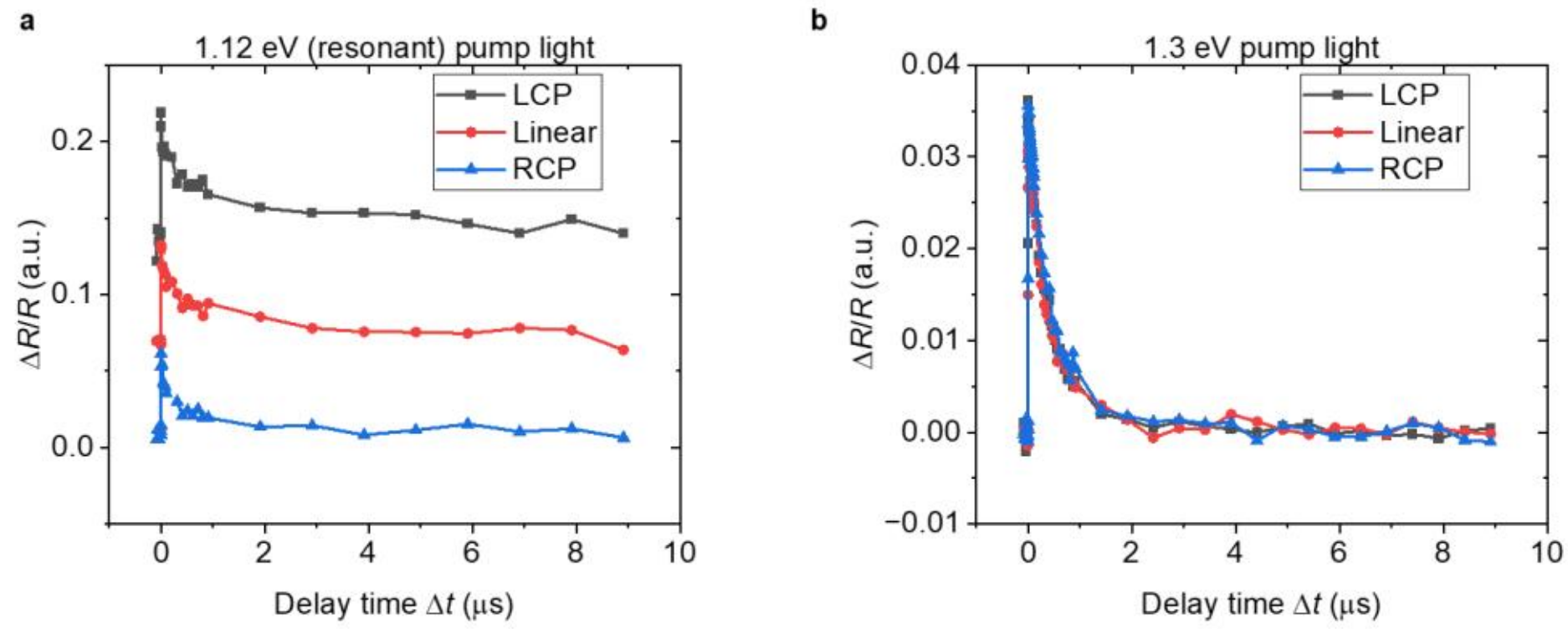

**Extended Data Fig. 3: Valley selection rules under resonant and non-resonant pumping. a**, Pump-induced spin-valley excitations measured using a 1.12 eV resonant pump light. The distinct signal amplitude and dynamics between LCP and RCP pump indicates much higher spin excitation density in the former case, as expected from well-defined valley selection rules. **b**, Same measurement as (**a**) using a 1.3 eV non-resonant pump light. The signal shows negligible dependence on the pump polarization (LCP, Linear, RCP), indicating no valley selection rules. Measurements are taken at $T = 2.5$ K, $\nu = -1$, $D = 6$ mV/nm with magnetic field $B_z = 100$ mT and pump fluence of 3 μJ/cm$^2$ in Device D1.

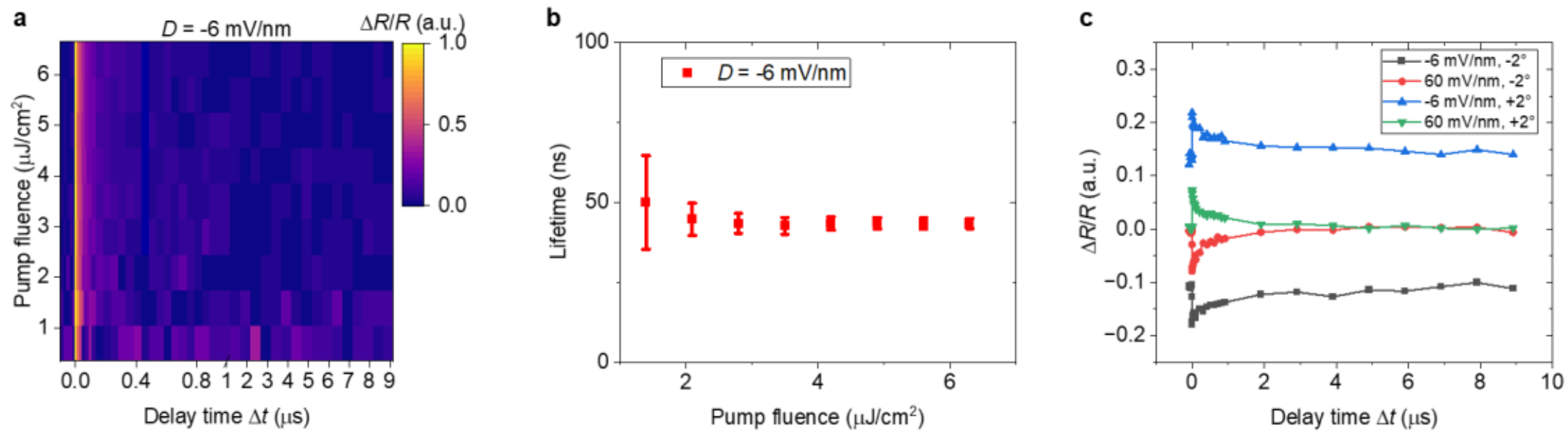

**Extended Data Fig. 4: Separation of charge (population) and spin-valley dynamics.** **a,** Pump fluence-dependent normalized dynamics measured using a parallel probe configuration to capture the electron-hole recombination process at $\nu$ = -1 and $D$ = -6 mV/nm (see Methods). **b,** Extracted charge dynamics shows a short, pump fluence-independent lifetime of < 100 ns. **c,** Spin-valley dynamics measured at $D$ = -6 mV/nm and $D$ = 60 mV/nm using near-cross probe configuration. The signals change signs when reversing the polarizer angle $\phi$, confirming their origin from pure spin-valley excitations (see Methods).

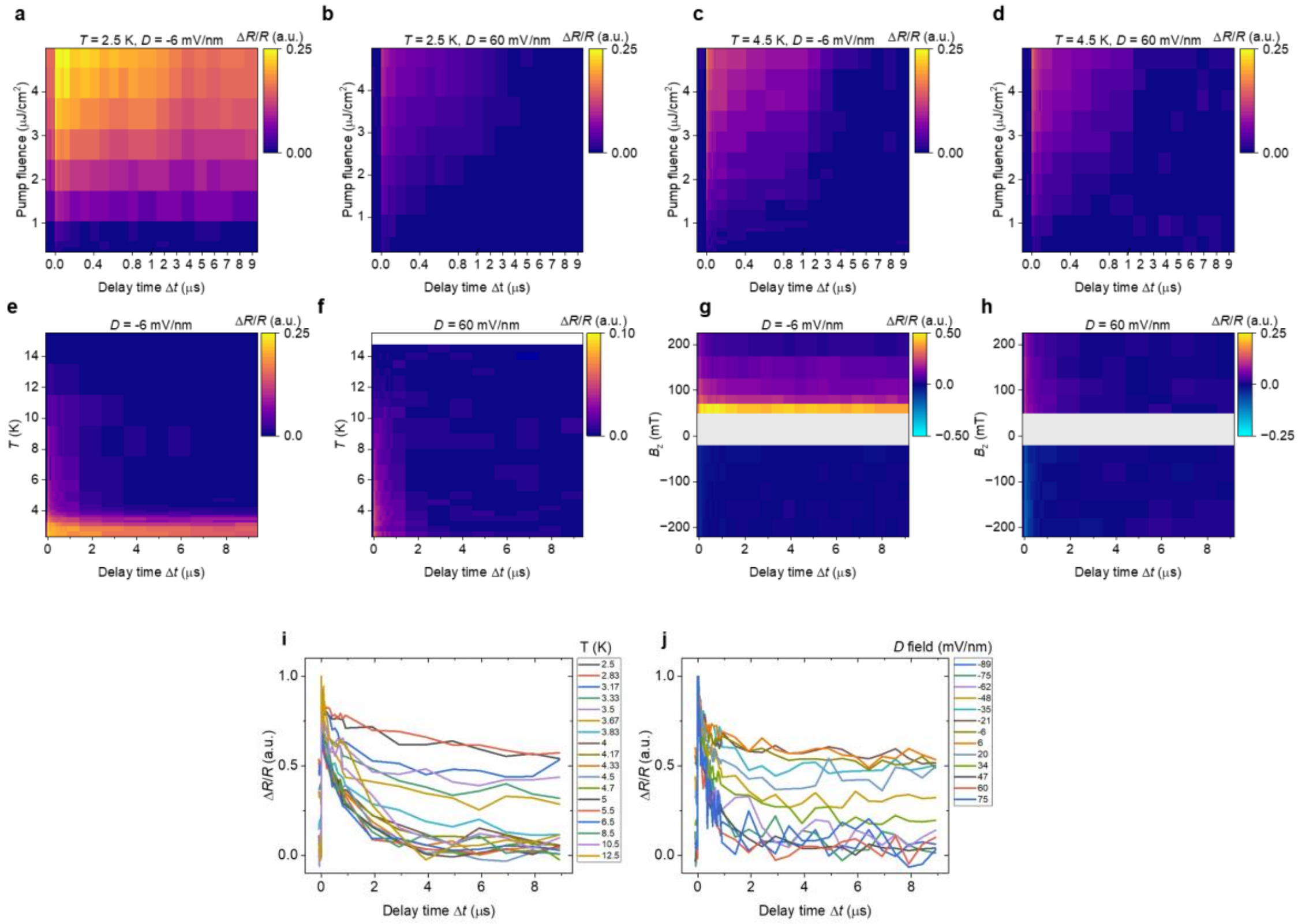

**Extended Data Fig. 5: Raw pump probe data. a-d**, unnormalized pump fluence-dependent spin-valley dynamics for main text Fig. 2a-d. **e,f,** Unnormalized temperature-dependent spin-valley dynamics for main text Fig. 3a and 3b. **g,h,** Unnormalized magnetic field-dependent spin-valley dynamics for main text Fig. 3e and 3f. **i,j,** Normalized delay-time traces at representative temperatures for $D$ = -6 mV/nm (**i**) and at representative $D$-fields for $T$ = 2.5 K (**j**).

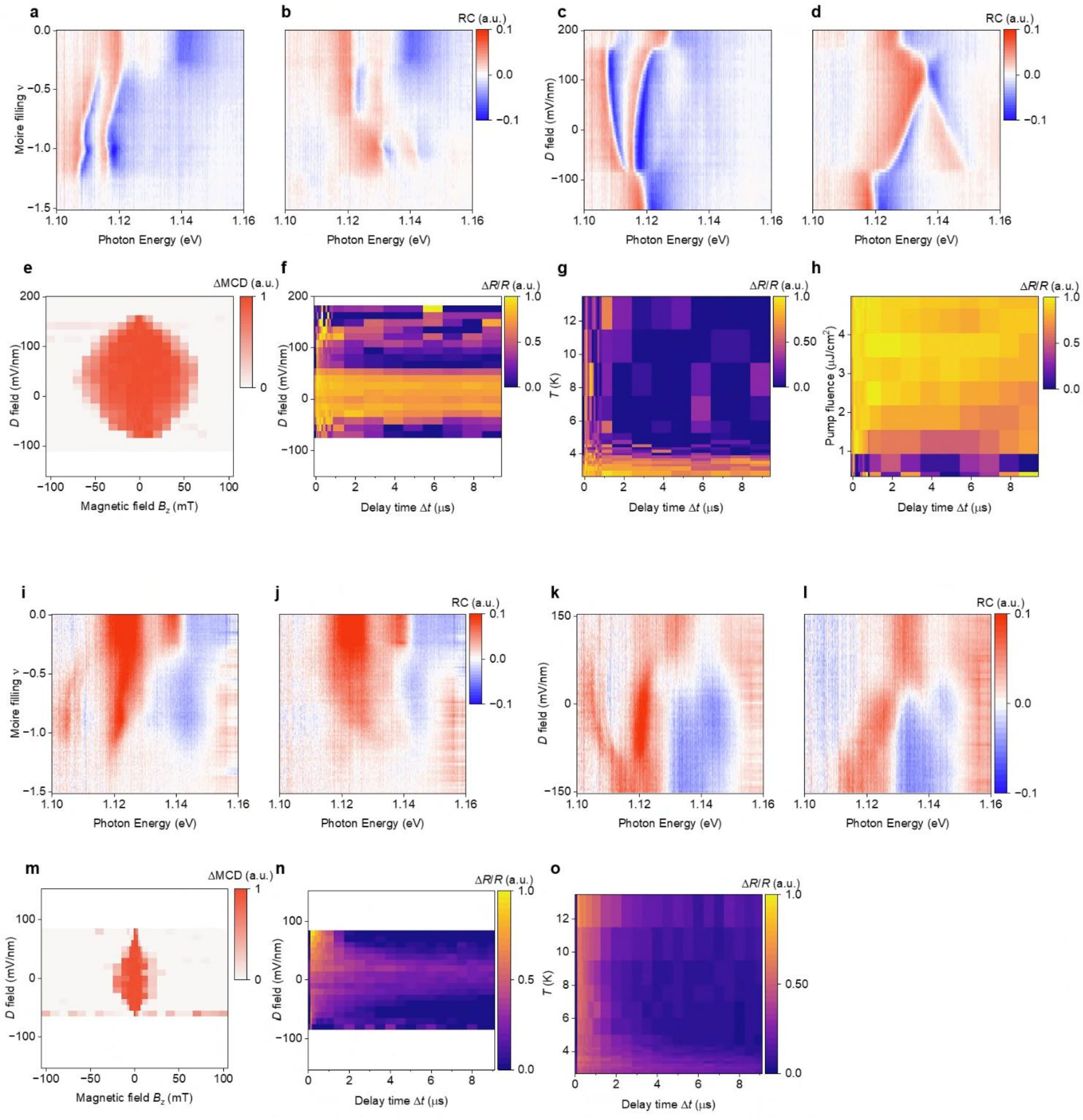

**Extended Data Fig. 6: Reproducibility of emergent spin-valley excitations in two additional devices, D2 and D3. a**,**b**, Polarization-resolved RC spectra of Device D2 (3.6° twist angle) as a function of moiré filling $\nu$ under LCP (**a**) and RCP (**b**) illumination. **c**,**d**, $D$-field-dependent RC spectra at moiré filling $\nu$ = -1 under LCP (c) and RCP (d) illumination. **e**, Magnetic hysteresis map showing the difference in MCD between forward and backward magnetic field sweeps at different $D$-fields. **f**, $D$-field-dependent normalized spin-valley dynamics at $\nu$ = -1 under $B_z$ = 100 mT. Long-lived metastable excitations emerge within a displacement field range narrower than the spin-valley polarized state in (**e**). **g**, Temperature-dependent spin-valley dynamics at $D$ = 0 mV/nm. The metastable component disappears around 3.8 K. **h**, Pump fluence-dependent spin-valley dynamics at $D$ = 0 mV/nm. The amplitude of metastable component shows threshold behavior with respect to pump fluence. These results confirm the observations from Device 1. All measurements are taken at base temperature $T$ = 2.5 K unless specified. **i-o,** Same as **a-g** for a third device D3 (3.7° twist angle).

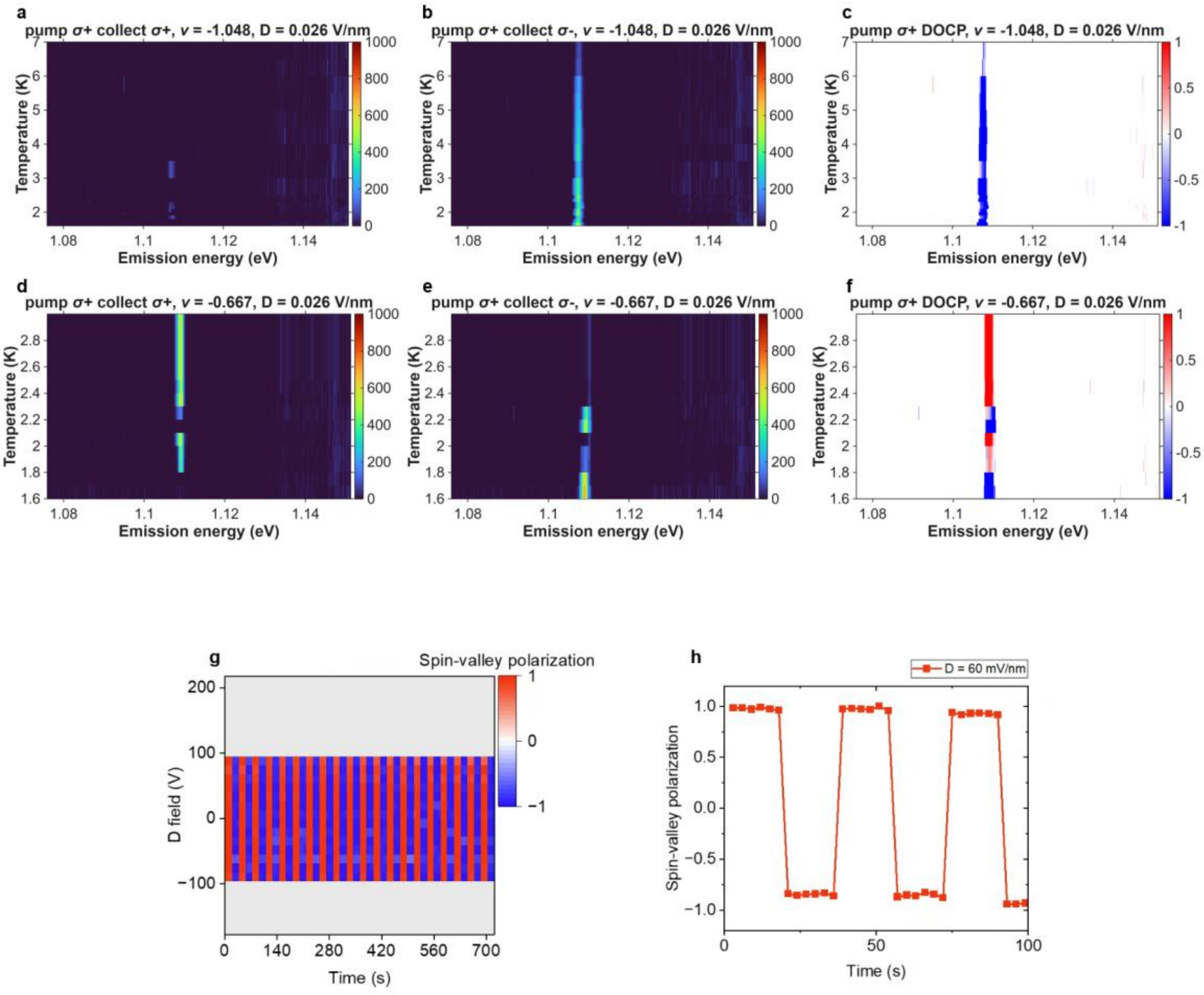

**Extended Data Fig. 7: Optical switching of the QAH and FQAH states in device D1. a-c**, Temperature-dependent, helicity-resolved PL spectra showing optical switching of the QAH state ($\nu$ = -1.048, $D$ = 0.026 V/nm). The panels correspond to $\sigma^+$ PL (**a**), $\sigma^-$ PL (**b**), and the degree of circular polarization (DOCP) (**c**) after switching the QAH state with $\sigma^+$ pump. The switching is reliable at lower and higher temperatures but becomes unstable between 3 and 3.5 K. **d-f**, Similar temperature-dependent switching measurements for the FQAH state ($\nu$ = -0.667, $D$ = 0.026 V/nm). The switching fails between 1.8 ~ 2.2 K but is reliable both below and above. The failed switching above 2.4 K is likely due to dominant heating effect that brings the sample above $T_c$. Measurements in a-f are performed in a cryostat with a base temperature of 1.6 K. **g,** Optical switching measurements of device D1 at $\nu$ = -1 and $T$ = 2.5 K, performed under the same configurations as the pump-probe measurements. The magnetization can be reliably switched at all $D$-fields and remain stable for at least 20 seconds after switching. **h,** Representative time trace at $D$ = 60 mV/nm.

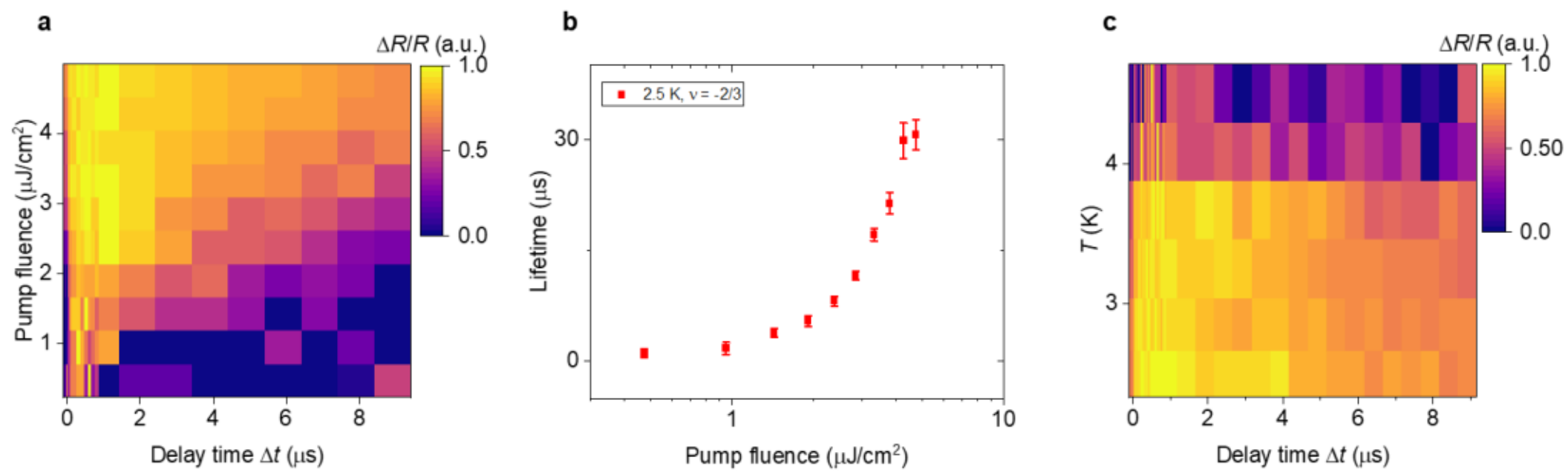

**Extended Data Fig. 8: Dynamics of the FQAH state. a**,**b**, Pump fluence-dependent normalized spin-valley dynamics (**a**) and the extracted lifetime (**b**) of the FQAH state ($\nu$ = -2/3) at $T$ = 2.5 K and $D$ = 0 mV/nm in Device D1. The relaxation lifetime exhibits a continuous increase with pump fluence, consistent with ordinary domain dynamics. **c**, Temperature-dependent normalized spin-valley dynamics at $\nu$ = -2/3 and $D$ = 0 mV/nm. No apparent change in dynamics is observed below $T_c$ due to the higher base temperature of the pump-probe measurements (2.5 K).

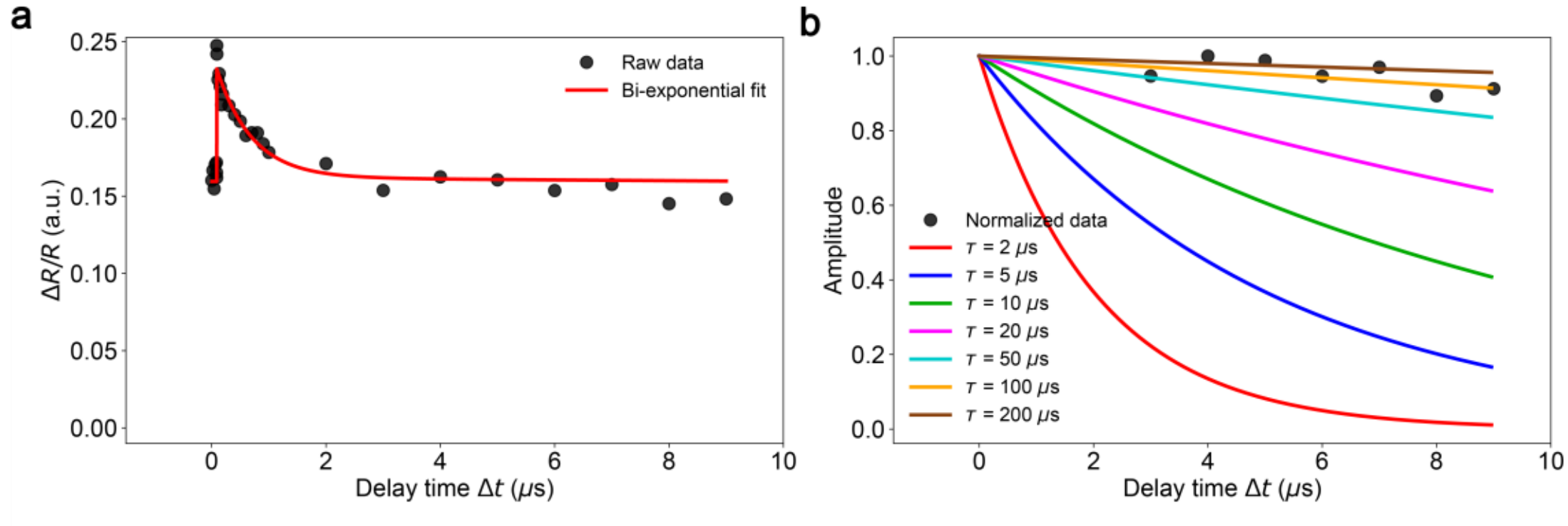

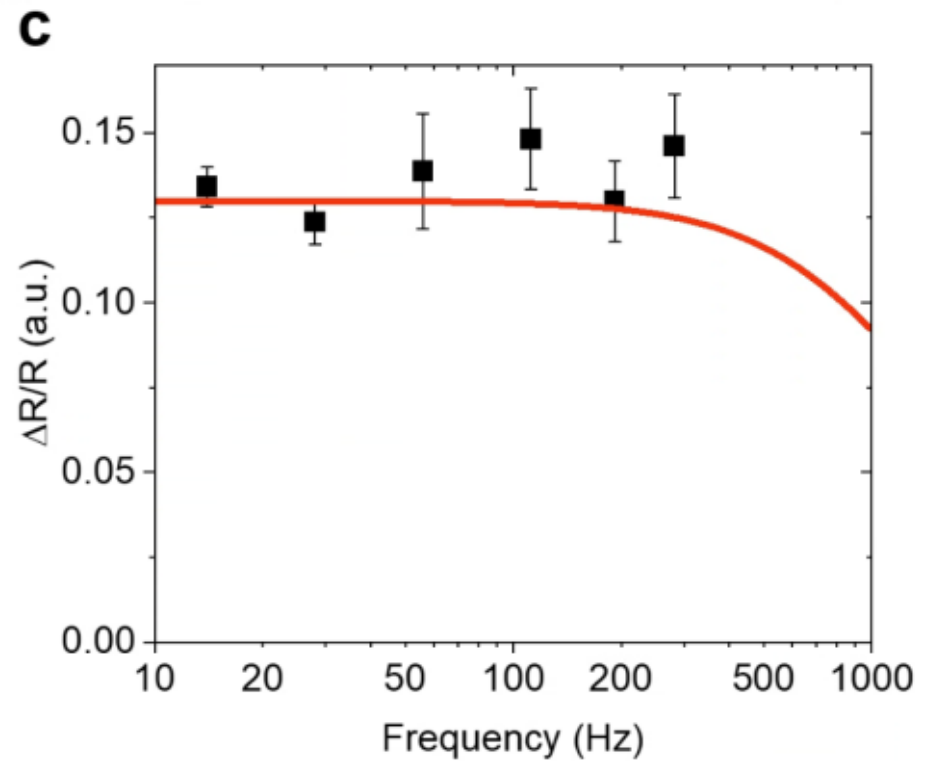

**Extended Data Fig. 9: Estimating lifetime of chiral domain walls. a,** Example bi-exponential fit (red line) of the measured spin-valley dynamics (black circles). **b,** Simulated dynamics of a single component with lifetime $\tau$ = 2, 5, 10, 20, 50, 100 and 200 μs, accumulated over $N$ = 50 pump pulses (see Methods). Black circles are the measured dynamics at $\Delta t > 2$ μs from **a**, indicating a lifetime of at least 50 μs. **c,** Pump-induced spin-valley polarization measured at a fixed delay $\Delta t$ = 9.99 μs as a function of the pump modulation (chopper) frequency, with all other conditions kept identical. The signal shows negligible decrease up to 300 Hz, indicating dynamics faster than 1 kHz. The red curve is the standard response of a 1 kHz bandwidth. Error bars represent the standard deviation of repeated measurements. Experimental data in all panels are taken at $\nu$ = -1, $D$ = -6 mV/nm, $T$ = 2.5 K, $B_z$ = 100 mT.

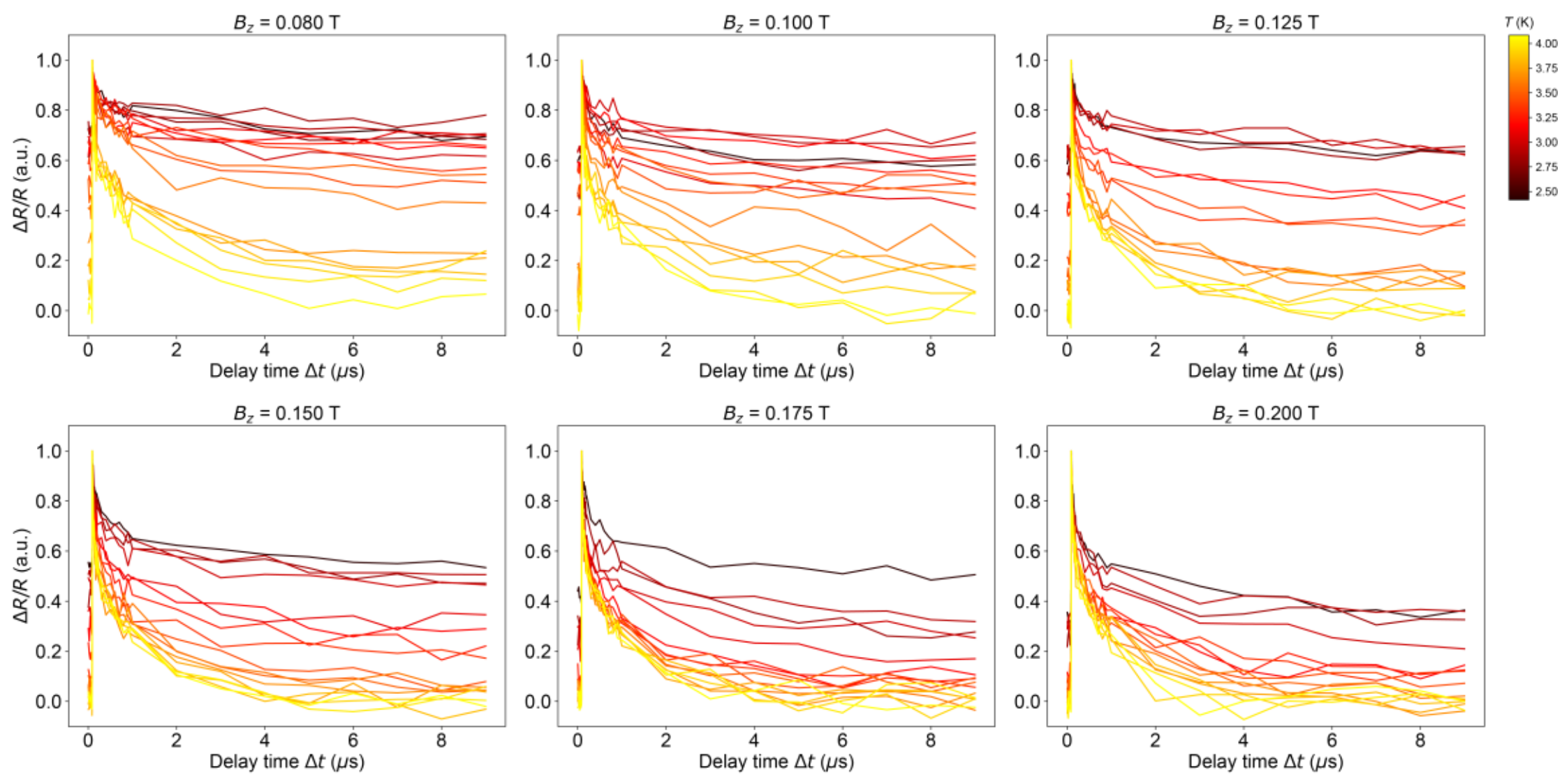

**Extended Data Fig. 10: Temperature dependent spin-valley dynamics at different *$B_z$*.** At each magnetic field, the onset temperature $T^*$ of the long-lived component is extracted and summarized in Fig. 4d.